# Roughness of Molecular Property Landscapes and Its Impact on Modellability

Matteo Aldeghi [1], David E. Graff [1,2], Nathan Frey [3], Joseph A. Morrone [4], Edward O. Pyzer-Knapp [5], Kirk E. Jordan [6], Connor W. Coley [1,7,*]

[1] Department of Chemical Engineering, Massachusetts Institute of Technology, Cambridge, MA 02139, USA
[2] Department of Chemistry and Chemical Biology, Harvard University, Cambridge, MA 02138, USA
[3] Lincoln Laboratory, Massachusetts Institute of Technology, Lexington, MA 02421, USA
[4] IBM Thomas J. Watson Research Center, Yorktown Heights, NY 10598, USA
[5] IBM Research Europe, Daresbury, UK
[6] IBM Thomas J. Watson Research Center, Cambridge, MA 02142, USA
[7] Department of Electrical Engineering and Computer Science, Massachusetts Institute of Technology, Cambridge, MA 02139, USA
[*] Email: ccoley@mit.edu

In molecular discovery and drug design, structure-property relationships and activity landscapes are often qualitatively or quantitatively analyzed to guide the navigation of chemical space. The roughness (or smoothness) of these molecular property landscapes is one of their most studied geometric attributes, as it can characterize the presence of activity cliffs, with rougher landscapes generally expected to pose tougher optimization challenges. Here, we introduce a general, quantitative measure for describing the roughness of molecular property landscapes. The proposed roughness index (ROGI) is loosely inspired by the concept of fractal dimension and strongly correlates with the out-of-sample error achieved by machine learning models on numerous regression tasks.

## Introduction

Structure-activity relationships (SARs) and activity landscapes are important concepts in cheminformatics and medicinal chemistry, and they are often used to guide the navigation of chemical space during molecular optimization campaigns (e.g., lead optimization)[1–3]. Quantitative SAR (QSAR) modeling uses numerical representations of chemical matter with machine learning (ML) models for the prediction of biological activity. QSAR concepts have been adopted more broadly across chemistry research through the application of structure-property relationships (SPR) and associated QSPR[4].

Roughness is one the most frequently discussed attributes of structure-property landscapes, perhaps owing to the interest in the identification of "activity cliffs" in drug design[5–8]. Activity cliffs are sharp changes in compound activity as a result of seemingly small structural changes, which can present a major obstacle in the development of accurate QSPR models[5,9–11]. As a result, a number of studies have focused on their identification[12–14] and prediction[15–17], typically by analyzing or predicting affinity differences in matched molecular pairs. It is clear that the presence (or absence) of activity cliffs is intrinsically linked to the roughness (or smoothness) of the property landscape. Smooth landscapes are generally favored because they lead to better interpretability, as well as predictability, by chemists; they are more easily modeled by ML algorithms; and they facilitate similarity-based virtual screening[18]. These benefits may thus affect strategic decisions during the discovery process, such as which compounds to prioritize for lead optimization.

Given the interest in quantitatively describing structure-property landscapes, different approaches have been developed to analyze their topography. To visualize property landscapes, Peltason et al.[19] have proposed to use multidimensional scaling to project high-dimensional representations onto the 2D plane and display SPRs as three-dimensional landscapes. These 3D landscapes have been combined with SPR matrices[20] and molecular grid maps to provide a tool for their organization and analysis[21]. Image analysis techniques have also been used to classify 3D property landscapes based on their degree of ruggedness[22], and to define a measure of similarity between them[23].

Among the indices developed to capture characteristics of property landscapes quantitatively, there is the Structure-Activity Landscape Index (SALI)[14]. SALI is a pairwise score that captures the magnitude of the property change with respect to distance of two compounds in chemical space,

$$\text{SALI}_{i,j} = \frac{|y_i - y_j|}{d(x_i, x_j)},$$

where $x$ is a numerical representation of a molecule, $y \in \mathbb{R}$ is its property, and $d$ is a distance metric. SALI effectively corresponds to the slope of a straight line connecting the points $(x_i, y_i)$ and $(x_j, y_j)$ in the metric space defined by $d$. The largest value in a dataset corresponds to the observed Lipschitz constant of the SPR function $f$ given the available data. Heatmaps and graph representations can be obtained from the full SALI matrix and can be used to identify the most significant property cliffs in the dataset. This index is not upper bounded, taking values between zero (when $|y_i - y_j| = 0$) and infinity (when $d(x_i, x_j)$ tends to zero).

As a global, rather than local, measure of roughness for a given molecular dataset, Peltason and Bajorath have proposed the SAR index (SARI)[13]. It is defined as the average of a continuity and a discontinuity scores,

$$\text{SARI} = 0.5 \times [\text{score}_{\text{cont}} - (1 - \text{score}_{\text{disc}})].$$

The continuity score is derived from the property-weighted average of pairwise compound similarity, while the discontinuity score is defined as the average potency difference between ligands with Tanimoto similarity greater than 0.6, multiplied by the pairwise ligand similarity[13]. SARI is conveniently defined between zero (rougher landscape) and one (smoother landscape). However, the raw scores are first standardized based on the mean and standard deviation of the scores obtained for a set of 16 reference datasets. A local version of SARI has also been developed, and has been used to create molecular networks to organize and display similarity and potency relationships within compound datasets[24].

Another quantitative measure that has been proposed to describe property landscapes is the modellability index (MODI)[25]. MODI tries to predict whether an accurate classification model is achievable, for a given training set, on the basis of the agreement/disagreement in label between nearest neighbor pairs,

$$\text{MODI} = \frac{1}{K} \sum_{i=1}^{K} \frac{N_i^{\text{same}}}{N_i},$$

where $K$ is the number of classes, $N_i$ is the number of compounds in each class, and $N_i^{\text{same}}$ is the number of compounds in each class having their nearest neighbor belonging to the same class. MODI is defined between zero and one; the more activity cliffs that are present, the closer to zero it is. The original formulation was later expanded by the same authors[26], as well as by Ruiz and Gómez-Nieto, who considered within- and between-class nearest neighbor pairs, as well as k-neighbors[27]. The approach was further generalized in order to be applied to regression tasks. Golbraikh *et al.*[26] did so by considering the performance of k-nearest neighbor models, and Ruiz and Gómez-Nieto by binarizing the dataset[28]. A different approach was instead taken by Marcou *et al.*[29], who use kernel-target alignment[30] as a measure of similarity between the descriptor and the property spaces.

Despite the development of the quantitative measures mentioned above, a truly general measure of roughness for molecular property landscapes is still missing. Being a local measure, SALI cannot capture the roughness of a molecular property landscape in a single scalar value. While SARI can do so, it relies on user-defined hyperparameters that need to be set heuristically, such as a similarity threshold for the discontinuity score, and reference datasets for standardization. Finally, MODI applies to primarily classification tasks, and extensions to regressions have been challenging. Contrary to SALI and SARI, MODI and related indices have been tested for their ability to anticipate the predictive performance of ML models, which is the primary evaluation approach we adopt in this work. There is a clear relationship between the geometry of structure-property landscapes and the ability of ML algorithms to model it, as observed already by Maggiora[9].

In this work, we propose a new measure of roughness for metric spaces that is directly applicable to molecular datasets. This Roughness Index (ROGI) captures the global ruggedness character of a normalized dataset as a single scalar value between zero and one, where zero corresponds to a flat surface and one to a surface in which all nearest neighbors display property values at the opposite extremes. Contrary to most of the approaches described above, it has no hyperparameters once a molecular representation and a metric have been defined. It naturally applies to regression tasks for any property of interest, as well as to binary classification tasks. To test the reliability and informativeness of ROGI, we evaluated its ability to anticipate the predictive performance of various ML models on a number of regression tasks, and found that it correlates with out-of-sample model error more reliably than existing indices.

# Methods

Chemical spaces, for example stable molecules at ambient temperature and pressure or more well-defined subsets like the set of drug-like molecules, can be defined as metric spaces, where each molecule $i$ is associated with a representation $x_i \in \mathcal{X}$, and a distance metric $d : \mathcal{X} \times \mathcal{X} \mapsto \mathbb{R}_{\geq 0}$ defines molecular dissimilarity. $d(\cdot, \cdot)$ is non-negative, symmetric, and generally satisfies the triangle inequality for most metrics used in practice. It may be the case that $d(x_i, x_j) = 0$ even if molecules $i$ and $j$ are distinct (e.g., binary fingerprints with finite radius or bit collisions), such that the space could be more appropriately described as pseudo-metric. Nevertheless, the above is typically all one can assume about a chemical space, which makes it challenging to define geometric properties, such as roughness, using measures that have been conceived for Euclidean spaces, like those used in topography, geology, and materials science[31–39]. The properties of molecules and materials we are interested to predict are described by continuous variables, as in regression tasks, with ML algorithms trying to model the underlying function $f$ that maps molecules to properties, $f : \mathcal{X} \mapsto \mathbb{R}$.

ROGI is loosely inspired by the concept of fractal dimension, which is an index of complexity comparing how some property of an object changes with the scale at which it is measured[40–43]. For example, by measuring the rate at which the observed coastline length increases as a function of a decreasing measuring unit (e.g., by using an increasingly short measuring stick), the roughness of a coastline can be quantified by its fractal dimension[40,44]. Essentially, an increasingly coarse-grained view of a certain object (e.g., a coastline) is taken, and the rate at which some of its properties change relates to the object's complexity. In the same vein, to describe the roughness of a molecular property landscape, we progressively coarse-grain a molecular dataset and observe how the dispersion of a molecular property of interest is affected.

## *Formulation of the roughness index*

The intuition behind the proposed approach is depicted in Figure 1. For this example, consider a dataset of molecules $\{x_i\}$ and associated property values $\{y_i\}$, where $x_i \in \mathcal{X}$ and $y_i \in \mathcal{Y} \subseteq \mathbb{R}$. Assume normalized, pairwise distances between all molecules in the dataset such that $d(x_i, x_j) \in [0, 1] \ \forall \ x_i, x_j \in \mathcal{X}$. We then cluster the dataset given different distance thresholds $t \in [0, 1]$ using complete-linkage clustering, such that the distance of any two elements in a cluster is at most $t$ (Figure 1a). Given $Y$ is a continuous property, we measure dispersion using the second central moment of its distribution, and more specifically we take its standard deviation $\sigma$. For every distance threshold $t$, we consider a dataset $\mathcal{D}_K^{(t)} = \{y_k, z_k\}_{k=1}^K$ where $K$ is the number of clusters, $y_k$ is the average molecular property within the cluster $k$, and $z_k$ is the cluster size. The weighted standard deviation, $\sigma_t$, of $\mathcal{D}_K^{(t)}$ is computed based on the weights $\{z_k\}_{k=1}^K$ (Figure 1b). This is equivalent to assigning the average property value $y_k$ to all members of each cluster and then computing the standard deviation for the whole dataset. At $t = 0$, each molecule belongs to its separate cluster, and $\sigma_0$ is the standard deviation of values in the original dataset. When $t = 1$, the dataset is described by a single cluster with zero standard deviation. At intermediate values of $t$, we effectively have a coarse-grained version of the dataset where each cluster $k$ is represented by a fictitious average molecule with an average property value $y_k$ (Figure 1a). $\sigma_t$ is guaranteed to decrease monotonically, from its original value $\sigma_0$ to zero, as $t$ goes from zero to one.

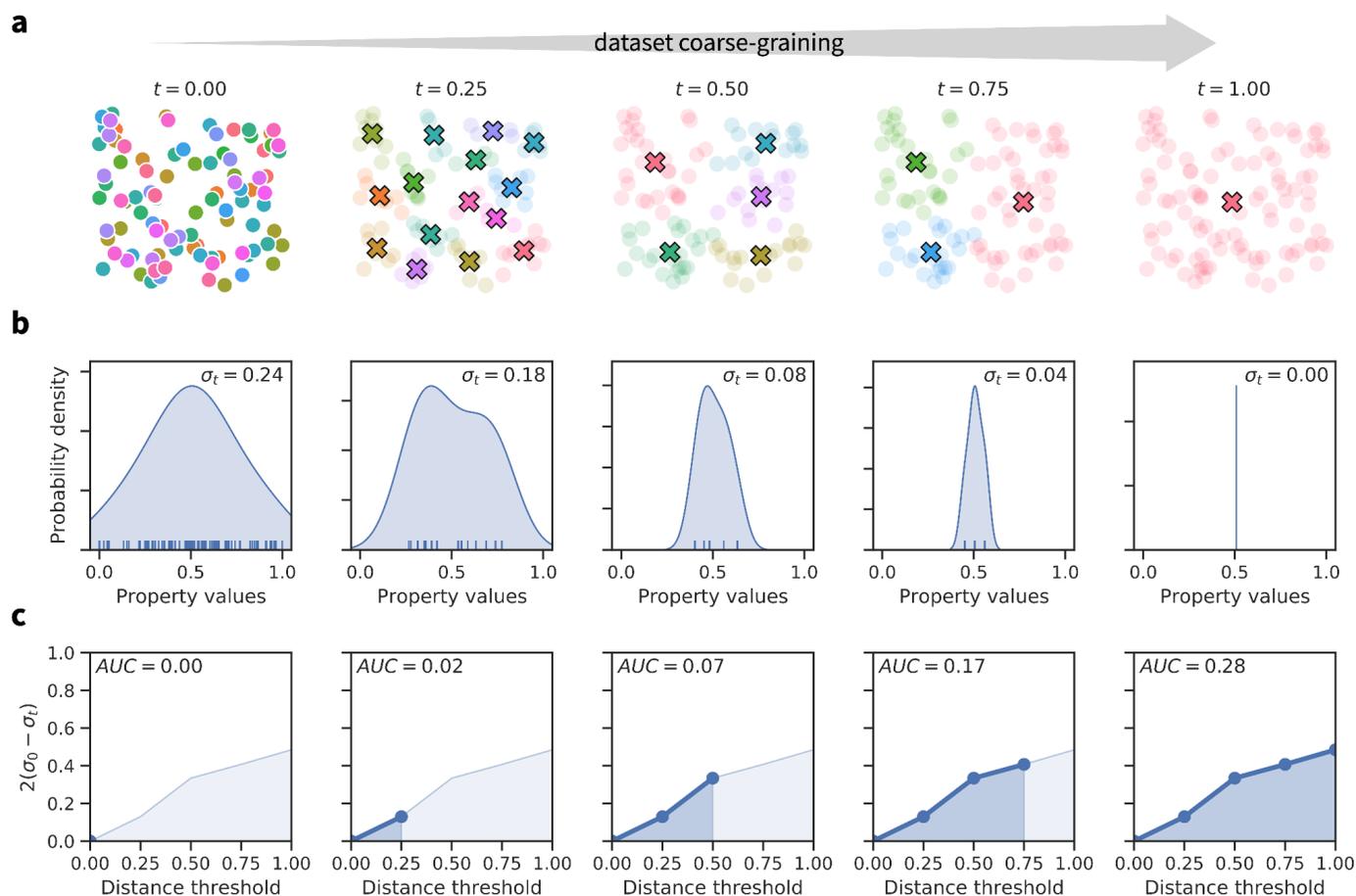

**Figure 1 | Schematic of the concept underlying the roughness index**. (a) Clustering of the dataset. The molecular dataset is clustered using different distance thresholds $t$ by complete-linkage, creating coarse-grained versions of the dataset where each cluster is represented by a hypothetical molecule with average property (crosses). (b) Decreasing property dispersion. The weighted standard deviation $\sigma_t$ captures the property dispersion at different levels of dataset coarse-graining (i.e., clustering at different $t$). (c) Pace of dispersion loss, measured as the area under the curve (AUC). The rate at which we lose property dispersion is measured by the area under the curve plotting dispersion loss as a function of clustering distance $t$. Dispersion loss is defined as $\sigma_0 - \sigma_t$, with the factor of 2 used for normalization.

As we coarse-grain the molecular dataset, we monitor the loss in dispersion $\sigma_0 - \sigma_t$ (Figure 1). Intuitively, if similar molecules have extremely different property values, they will be clustered at low $t$ values and the dispersion across clusters will decrease rapidly. Conversely, if similar molecules have similar property values, replacing these by their average will have a small effect on the overall dispersion of the property across clusters, such that $\sigma_0 - \sigma_t$ will increase slowly. To measure how quickly dispersion is lost as $t$ increases we integrate $\sigma_0 - \sigma_t$ between zero and one. For normalized property values and pairwise distances, ROGI is thus defined as

$$\mathrm{ROGI} = \int_0^1 2(\sigma_0 - \sigma_t)\,\mathrm{d}t,$$

where $\sigma_0$ and $\sigma_t$ are here the standard deviations obtained from normalized property values (Figure 1c). Note that, while the ROGI was primarily devised for regression, it may also be applied to binary classification as is. In the future, expanding ROGI to multi-class classification may also be possible by considering, e.g., information entropy as a measure of dispersion.

The above is equivalent to computing ROGI for the original dataset, before scaling it according to the largest molecular distance and property ranges,

$$\frac{2}{(\max(\mathcal{Y}) - \min(\mathcal{Y})) \cdot \max(d_\mathcal{X})} \int_0^{\max(d_\mathcal{X})} \sigma_0 - \sigma_t \, dt,$$

where $\max(d_\mathcal{X})$ is the largest distance between any two elements in $\mathcal{X}$. The term before the integral normalizes the index between zero and one. In fact, given any set of property values, the largest standard deviation achievable is $0.5(\max(\mathcal{Y}) - \min(\mathcal{Y}))$. And given that $\max(d_\mathcal{X})$ is the largest distance attainable based on the chosen representation and metric, it also the largest value of $t$, for which only one cluster exists. When the metric used is the Jaccard distance between binary fingerprints (i.e., $1 - T_C$, where $T_C$ is the Tanimoto similarity widely used for structure comparison[45,46]), $d(x_i, x_j) \in [0,1] \; \forall \; x_i, x_j \in \mathcal{X}$, such that distances are already normalized. When $d(x_i, x_j) \in [0, \infty)$, such as for p-norm distances when using descriptors, $\max(d_\mathcal{X})$ is the largest distance in principle attainable between any two molecules in chemical space. As this information is usually not available, $\max(d_\mathcal{X})$ can be approximated by the largest distance within the hyper-rectangle defined by the descriptor values.

For any dataset, the above integral can be approximated numerically according to the available resolution of $t$, as only a finite set of pairwise distances is available, which poses a limit on how often the clusters of the dataset change. In addition, distances are unlikely to be uniformly distributed. In our implementation, we use the trapezoidal rule (Figure 1c) with $\Delta t = 0.01 \cdot \max(d_x)$. The number of potential clustering thresholds $t$ grows quadratically with dataset size, so bounding $dt$ can significantly reduce the cost of computing the ROGI for large datasets without losing much accuracy. Overall, the computation of the ROGI inherits the quadratic scaling of hierarchical clustering, $\mathcal{O}(KN^2)$, where $K$ is the number of clusters and $N$ is the number of elements in the molecular dataset.

## Datasets and data analysis

*Evaluation of the roughness index.* As there is no unambiguous definition or ground-truth value for molecular dataset roughness, we rely on the connection between the roughness of a property landscape and the ability of a ML model to accurately model it. Given a certain structure-property landscape, the more data and/or the more expressive the ML model used, the more accurate predictions one should be able to achieve in a random cross-validation. Given the same amount of data available and the same ML model, smoother landscapes should be more easily modeled than rougher ones. We thus evaluated the ability of ROGI to predict out-of-sample model error on a variety of datasets. For regression tasks, we examine the relative, i.e., normalized, root-mean-square error (RMSE) in a random cross validation. For classification, we examine the correlation between the ROGI and binary accuracy.

*Toy datasets*. Six two-dimensional analytical functions (F1 to F6), for which roughness can be qualitatively assessed visually, were used to validate the ROGI approach. Datasets were created by sampling uniformly from the domain [0,1]$^2$ of these functions. Details about these analytical functions and their implementation are provided at https://github.com/coleygroup/rogi-results.

*Chemistry datasets.* Three sets of regression tasks were used in this work. Structure-property landscapes related to regression tasks were retrieved from the Therapeutic Data Commons (TDC)[47] using the Python library *PyTDC* (v. 0.3.6), and from the previous work of van Tilborg *et al.*[11] A total of 55 regression datasets, split across three groups, were considered. (1) The group of datasets referred to as "ZINC+GuacaMol" comprised 13 datasets with 2000 molecules randomly sampled from ZINC[48], for which the properties were computed via the following GuacaMol[49] oracles: QED[50], LogP[51], Celecoxib_Rediscovery[49], Aripiprazole_Similarity[49], Median 1[49], Osimertinib_MPO[49], Fexofenadine_MPO[49], Ranolazine_MPO[49], Perindopril_MPO[49], Amlodipine_MPO[49], Zaleplon_MPO[49], Valsartan_SMARTS[49], Scaffold Hop[49]. (2) The second group of datasets, referred to as "TDC", comprised 12 datasets that featured pharmaco-kinetic and toxicological properties, and were obtained from the TDC: Caco2_Wang[52], Lipophilicity_AstraZeneca[53,54], Solubility_AqSolDB[55], HydrationFreeEnergy_FreeSolv[56], PPBR_AZ[54], VDss_Lombardo[57], Half_Life_Obach[58], Clearance_Hepatocyte_AZ[54,59], Clearance_Microsome_AZ[54,59], LD50_Zhu[60], herg_central/hERG_at_1uM[61], herg_central/hERG_at_10uM[61]. (3) The third group of datasets, referred to as "ChEMBL", comprised 30 SAR datasets from ChEMBL[62] that were curated by van

Tilborg et al.[11]. To reduce the computational cost of performing these tests, dataset sizes were capped at 10,000 molecules; datasets containing a larger number of entries were subsampled at random (using a fixed seed for reproducibility).

50 pharmaco-kinetic and toxicological datasets associated with binary classification tasks were also taken from the TDC[47]. These were HIA_Hou[63], Pgp_Broccatelli[64], Bioavailability_Ma[65], BBB_Martins[66], CYP2C19_Veith[67], CYP2D6_Veith[67], CYP3A4_Veith[67], CYP1A2_Veith[67], CYP2C9_Veith[67], CYP2C9_Substrate_CarbonMangels[68], CYP2D6_Substrate_CarbonMangels[68], CYP3A4_Substrate_CarbonMangels[68], hERG[69], AMES[70], DILI[71], Skin Reaction[72], Carcinogens_Lagunin[73,74], ClinTox[75], herg_central/hERG_inhib[61], all 12 datasets from the Tox21 challenge[76], and 19 datasets from ToxCast[77] selected reproducibly at random. As with regression tasks, dataset sizes were capped at 10,000 molecules.

*Machine learning models.* The above datasets were modeled with a range of baseline ML algorithms available in *scikit-learn* (v1.1.1)[78]. We selected five approaches to cover nearest neighbor, linear, tree-based, kernel, and deep learning methods. More specifically, for regression we used k-nearest neighbor (KNN) regression, partial least squares (PLS) regression, random forest (RF) regression, support vector regression (SVR), and a multi-layer perceptron (MLP). Similarly, we used KNN classification, logistic regression (LR), RF classification, support vector classification (SVC), and an MLP for classification tasks. In all cases, we used the default hyperparameters in *scikit-learn*, with the exception of RF for which we used 50 trees.

*Chemical representations*. Molecules were represented either by fingerprints or a set of descriptors. We used Morgan binary fingerprints as implemented in *RDKit*[79] (v2022.03), with 2048 bits and radius 2. As descriptors, we chose a set of 16 physico-chemical properties generally applicable across tasks: molecular weight, fraction of $sp^3$ carbons, number of hydrogen bonds acceptors and donors, number of nitrogen and oxygen atoms, number of NH and OH groups, number of aliphatic and aromatic rings, number of aliphatic and aromatic heterocycles, number of rotatable bonds, total polar surface area, LogP[51], and QED[50]. The descriptors chosen were not meant to be an ideal molecular representation for all prediction tasks studied, but simply a hypothetical one.

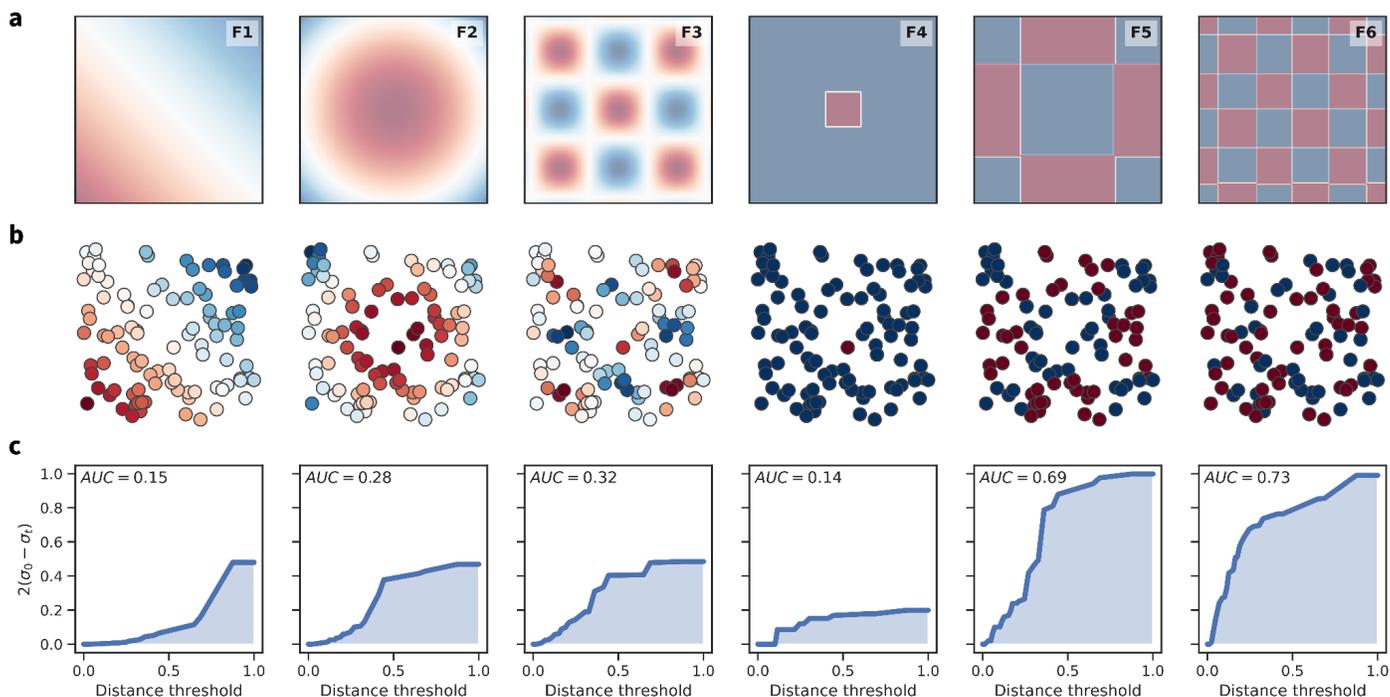

**Figure 2 | Toy surfaces and datasets illustrate the behavior of the roughness index**. (a) Two-dimensional surfaces with varying degrees of roughness. (b) Datasets of 100 samples drawn uniformly at random from the surfaces. (c) Area under the curve that defines the roughness index for each dataset.

# Results

## Validation on toy datasets

A set of six toy surfaces are used to demonstrate the behavior of ROGI (Figure 2a). Intuitively, roughness should increase between the continuous surfaces F1 and F3, and between the binary surfaces F4 to F6. To test this, we draw a uniform sample of size 100 from these two-dimensional surfaces, simulating a discrete molecular dataset (Figure 2b). Clustering was performed based on pairwise Euclidean distances, and ROGI was computed as the area under the $t$ vs. $2(\sigma_0 - \sigma_t)$ curve (Figure 2c), as described above. The roughness ranking suggested by the ROGI values closely follows the intuition one might have from visual inspection. Properties that change slowly with respect to the input representation are smoother than those that tend to change more abruptly, and the more minima and maxima are present in the landscape, the rougher it tends to be (larger ROGI values). This trend is especially noticeable for the surfaces with binary property values. These surfaces display sharp cliffs, and some are highly multimodal resulting in the presence of many cliffs. The larger the cliff area in the property landscape, the rougher it is, as reflected by their ROGI values. Note that extreme ROGI values are obtained for flat landscapes (ROGI of zero), and for landscapes in which all nearest neighbor pairs have opposite property values (ROGI of one; Figure S1).

Figure 3 shows parity plots that compare the average ROGI to the average relative, i.e., normalized, root-mean-square error (RMSE) obtained with a RF model for datasets of different sizes, sampled 50 times uniformly from the surfaces in Figure 2a. ROGI was able to correctly rank the difficulty of each regression task, with linear and monotonic correlations (Pearson correlation coefficients >0.9), with F1–F4 being deemed considerably less challenging than F5 and F6. A trend where the slope of the line of best fit decreases with increasing dataset size is visible. This effect seems logical. The ROGI of a property landscape has a well-defined value, even though the ROGI computed from a small sample is only an estimate (Figure S2). However, the accuracy of a model should increase (hence RMSE decrease) with increasing training set size. The effect of dataset size is thus a potentially confounding factor when comparing the roughness of different properties using different datasets, as we later demonstrate for datasets of binary molecular properties (Figure 5).

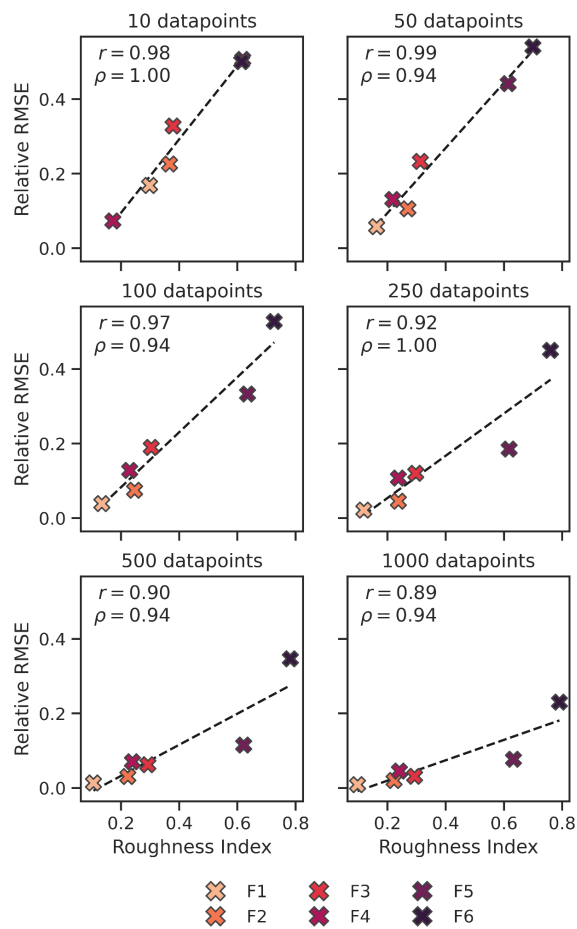

**Figure 3 | Correlation between the roughness index and cross-validated model error for toy datasets of different sizes.** The performance of a random forest model was evaluated as the relative (normalized) RMSE, using 10-fold cross validation. Datasets of different sizes (10, 50, 100, 250, 500, 1000) were sampled from the surfaces in Figure 2a. For each dataset size, 50 datasets were sampled, and both the relative RMSE and the ROGI were computed. The parity plots display the average relative RMSE against the ROGI across the 50 sample sets for F1–F6 and for the different dataset sizes tested.

## Roughness of continuous molecular properties

As done above for the toy surfaces, we evaluated the ability of ROGI to capture the roughness of a property landscape by testing its correlation with cross-validated model error. Here, we do so on a suite of 55 regression tasks (Methods). Because ROGI is model-independent, we performed these tests using five different ML algorithms, covering five different classes of supervised learning methods: k-nearest neighbors (KNN), linear models (partial least squares, PLS, for regression; logistic regression, LR, for classification), random forest (RF), support vector machines (SVR for regression; SVC for classification), and deep learning (multi-layer perceptron, MLP). Molecules in all datasets were represented either via Morgan fingerprints or a set of physico-chemical descriptors, as described in the Methods section.

ROGI positively, and often strongly, correlates with the predictive error of all regression ML models tested (Figure 4). This is particularly true for the datasets based on a randomly sampled subset of ZINC+GuacaMol and the pharmaco-kinetic and toxicology datasets from the TDC [47], both for fingerprint and descriptor representations. With the exception of ZINC+GuacaMol with RF and descriptors, correlations between ROGI values and model errors are above 0.8, and typically around or above 0.9. While we expected the range of dataset sizes (from 642 to 10,000) for the TDC tasks to worsen the correlation due to the size-dependent performance of ML models, correlations for the TDC datasets were generally strong (minimum correlation of 0.88; Figure 4).

The ZINC+GuacaMol dataset is particularly informative because it considers the same set of 2000 molecules but different properties thereof. For this group of datasets, the only case in which we did not observe a correlation between ROGI and cross-validated model error was when a RF model was used with molecules represented via descriptors. In this case, RF was able to predict the smoother and rougher properties with a similar degree of accuracy. These properties are defined as a combination of physico-chemical descriptors. As these descriptors were also used as input for the regression tasks, much of the roughness may be due to the presence of uninformative descriptors among informative ones, which RF was able to filter out more efficiently than other ML models.

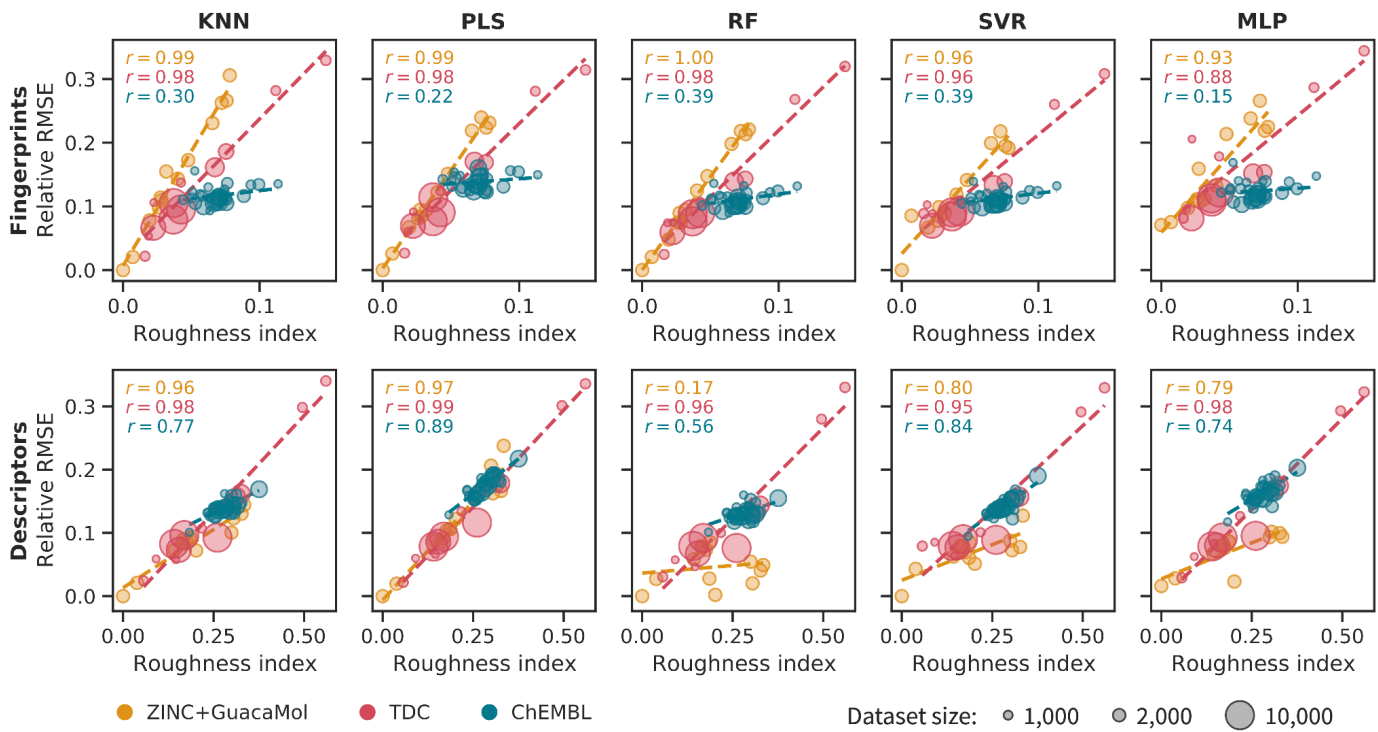

**Figure 4 | Correlation between the roughness index and cross-validated model error for a set of 55 regression tasks.** ZINC+GuacaMol refers to 13 datasets of 2000 molecules built from ZINC, and for which properties were computed using a set of oracle functions. TDC refers to 12 datasets used as regression benchmarks and provided by the Therapeutic Data Commons [47]. ChEMBL refers to 30 datasets curated from the ChEMBL database[62] by van Tilborg et al.[11]. The relative RMSE is the average, normalized RMSE obtained from 5-fold cross validation. Details of the fingerprints and descriptors used as molecular representations can be found in the Methods. In each plot, the Pearson correlation coefficient (*r*) between roughness and model error is reported. KNN = k-nearest neighbor regression; PLS = partial least square regression; RF = random forest; SVR = support vector regression; MLP = multi-layer perceptron.

For the ChEMBL datasets, ROGI displayed moderate-to-strong ($r$ = 0.56–0.89) correlations with model error when representing molecules with physico-chemical descriptors, but only weak correlations ($r$ = 0.15–0.39) when representing molecules with fingerprints. One possible reason for the lower correlations observed for the ChEMBL dataset is the much smaller range of both ROGI values and model errors obtained (Figure S4). The smaller the range of RMSEs, the more accurate ROGI estimates need to be to linearly correlate with model error. When using fingerprints, the KNN model returned RMSEs between 0.09 and 0.16 for ChEMBL, between 0.02 and 0.33 for TDC, and between 0.00 and 0.31 for ZINC+GuacaMol. Similar RMSE ranges were observed when using descriptors, yet the tight distribution of ROGI values associated with fingerprints might have exacerbated the issue (ROGI values of 0.04–0.11 for fingerprints, 0.18–0.38 for descriptors)

ROGI values for fingerprint representations were observed to be smaller, and in a tighter range, than those obtained for descriptors across all datasets. This effect is caused by a different distribution of pairwise distances. Tanimoto distances between molecules described with fingerprints are generally larger than distances obtained with the Euclidean metric applied to descriptors (Figure S5). Given the same set of molecules and property values, smaller distances between molecules imply a rougher surface, which is captured by ROGI. With larger distances between molecules instead, the ROGI estimate will tend toward lower values suggesting a smoother surface. In the limit of all molecules being maximally distant from each other (i.e., all pairwise distances equal to one), ROGI will be equal to zero, regardless of how the property values are distributed across molecules. In this case, however, a ROGI value of zero indicates a lack of sufficient information to assess the roughness of the structure-property landscape, as opposed to being evidence of a smooth surface. While these are extreme scenarios, it is important to interpret the ROGI value in the context of the dataset and distribution of distances between molecules.

As a comparison to existing approaches, the correlation of SARI to model error was evaluated on the ChEMBL dataset. Only this dataset was considered because SARI was developed specifically for protein-ligand binding affinity. We generally found significantly lower correlation between SARI and model error than were observed for ROGI (Figure S6). We also performed the same analysis, on all datasets, with the regression MODI (RMODI) described by Ruiz and Gómez-Nieto[28]. RMODI returned correlations with model errors higher than those obtained with SARI. However, with the exception of 5/30 instances (4 for ChEMBL with fingerprints, 1 for ZINC+GuacaMol with descriptors and RF), ROGI consistently provided stronger correlations (Table 1 and Figure S7). To reiterate an earlier observation, ChEMBL with fingerprints exhibits a narrow range of both ROGI scores and model RMSEs, which worsened the quantitative correlation, suggesting that consideration of multiple metrics in a consensus approach may be beneficial.

**Table 1 | Comparison between ROGI and RMODI on regression tasks.** Pearson correlation coefficients between ROGI or RMODI and cross-validated model error for a set of 55 regression tasks, spread across three groups (ZINC+GuacaMol, TDC, ChEMBL), are shown. Instances in which ROGI (blue) and RMODI (red) provided higher correlations are bolded.

| | | | KNN | PLS | RF | SVR | MLP |
|---|---|---|---|---|---|---|---|
| **Fingerprints** | ZINC+GuacaMol | ROGI | **0.99** | **0.99** | **1.00** | **0.96** | **0.93** |
| | | RMODI | 0.73 | 0.71 | 0.70 | 0.64 | 0.67 |
| | TDC | ROGI | **0.98** | **0.98** | **0.98** | **0.96** | **0.88** |
| | | RMODI | 0.32 | 0.30 | 0.27 | 0.12 | 0.24 |
| | ChEMBL | ROGI | 0.30 | **0.22** | 0.39 | 0.39 | 0.15 |
| | | RMODI | **0.60** | -0.04 | **0.58** | **0.62** | **0.54** |
| **Descriptors** | ZINC+GuacaMol | ROGI | **0.77** | **0.89** | 0.17 | **0.80** | **0.79** |
| | | RMODI | 0.49 | 0.20 | **0.72** | 0.46 | 0.59 |
| | TDC | ROGI | **0.98** | **0.99** | **0.96** | **0.95** | **0.98** |
| | | RMODI | 0.40 | 0.39 | 0.40 | 0.29 | 0.37 |
| | ChEMBL | ROGI | **0.77** | **0.89** | **0.56** | **0.84** | **0.74** |
| | | RMODI | 0.45 | -0.05 | 0.53 | 0.39 | 0.42 |

## Roughness of binary molecular properties

While ROGI was developed primarily for continuous structure-property landscapes (regression), it may be applied as is to discontinuous ones (e.g., binary classification). Figure 5 shows the correlation between ROGI values and the binary accuracy of classification models. These correlations were above 0.6 for all ML models using fingerprints as input, and above 0.8 for all models using descriptors (Figure 5). Here too, the datasets considered had a wide range of sizes, from 280 to 10,000. In this case, dataset size is a confounding factor. Most of the outliers—datasets that returned lower accuracy than expected given their ROGI value—were due to the smallest datasets (small circle markers in Figure 5).

The same analysis was performed with MODI, an established index for binary classification. While virtually no correlation between MODI and binary accuracy was observed (Table 2 and Figure S8), strong correlations were observed when balanced accuracy was used as the performance measure (Table 2 and Figure S9). The opposite was observed for ROGI, which is negatively correlated with balanced accuracy (Table 2 and Figure S10). These observations are consistent with how the two indices are defined. ROGI is a measure of global roughness and weights all instances equally, leading to lower values for more imbalanced datasets. (e.g., S4 in Figure 2). This bias is in line with how binary accuracy is defined. On the other hand, MODI considers all instances of the positive and negative class separately, takes the fraction of their nearest neighbors being in the same class, and averages them; thus effectively upweighting the importance of the minority class. Balanced accuracy similarly upweights the minority class by averaging true positive and true negative rates.

In summary, on balanced datasets (Figure S11) or for information retrieval tasks on imbalanced datasets, MODI is likely to be a better index of modellability. If the analysis of roughness or the overall fraction of correctly classified instances is instead the main goal, ROGI is likely to be more suitable.

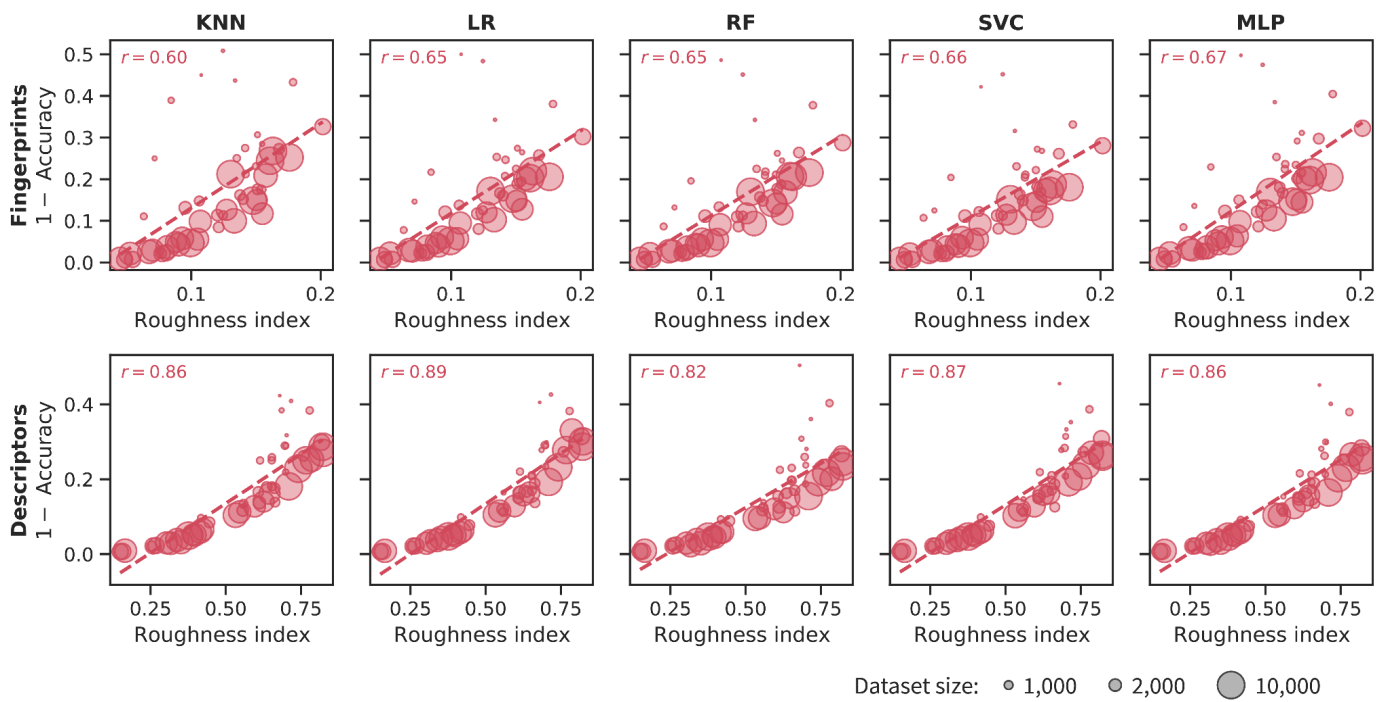

**Figure 5 | Correlation between the roughness index and cross-validated model error for a set of 50 classification tasks.** Accuracy refers to the average binary accuracy (i.e., the fraction of correctly classified labels) from 5-fold cross validation. In each plot, $r$ refers to the Pearson correlation coefficient, and $\rho$ to the Spearman correlation coefficient. Details of the fingerprints and descriptors used as molecular representations can be found in the Methods. KNN = k-nearest neighbor classification; LR = logistic regression; RF = random forest; SVC = support vector classification; MLP = multi-layer perceptron.

**Table 2 | Comparison between ROGI and MODI on binary classification tasks.** Pearson correlation coefficients between ROGI or MODI and cross-validated model error for a set of 50 classification tasks. Instances in which ROGI (blue) and MODI (red) provided higher correlations are bolded. The results show how their relative performance is highly dependent on whether accuracy or balanced accuracy is used as the measure of model performance. Because higher ROGI is meant to imply higher error, we calculate its correlation coefficient using one minus (balanced) accuracy.

|  | Evaluation Metric |  | KNN | PLS | RF | SVR | MLP |
|---|---|---|---|---|---|---|---|
| **Fingerprints** | Accuracy | ROGI | **0.60** | **0.65** | **0.66** | **0.66** | **0.67** |
|  |  | MODI | -0.03 | 0.07 | 0.07 | 0.06 | 0.08 |
|  | Balanced Accuracy | ROGI | -0.10 | -0.19 | -0.18 | -0.20 | -0.13 |
|  |  | MODI | **0.87** | **0.95** | **0.91** | **0.84** | **0.96** |
| **Descriptors** | Accuracy | ROGI | **0.86** | **0.89** | **0.82** | **0.87** | **0.86** |
|  |  | MODI | -0.15 | -0.25 | -0.08 | -0.19 | -0.18 |
|  | Balanced Accuracy | ROGI | -0.50 | -0.50 | -0.52 | -0.49 | -0.49 |
|  |  | MODI | **0.94** | **0.76** | **0.94** | **0.76** | **0.88** |

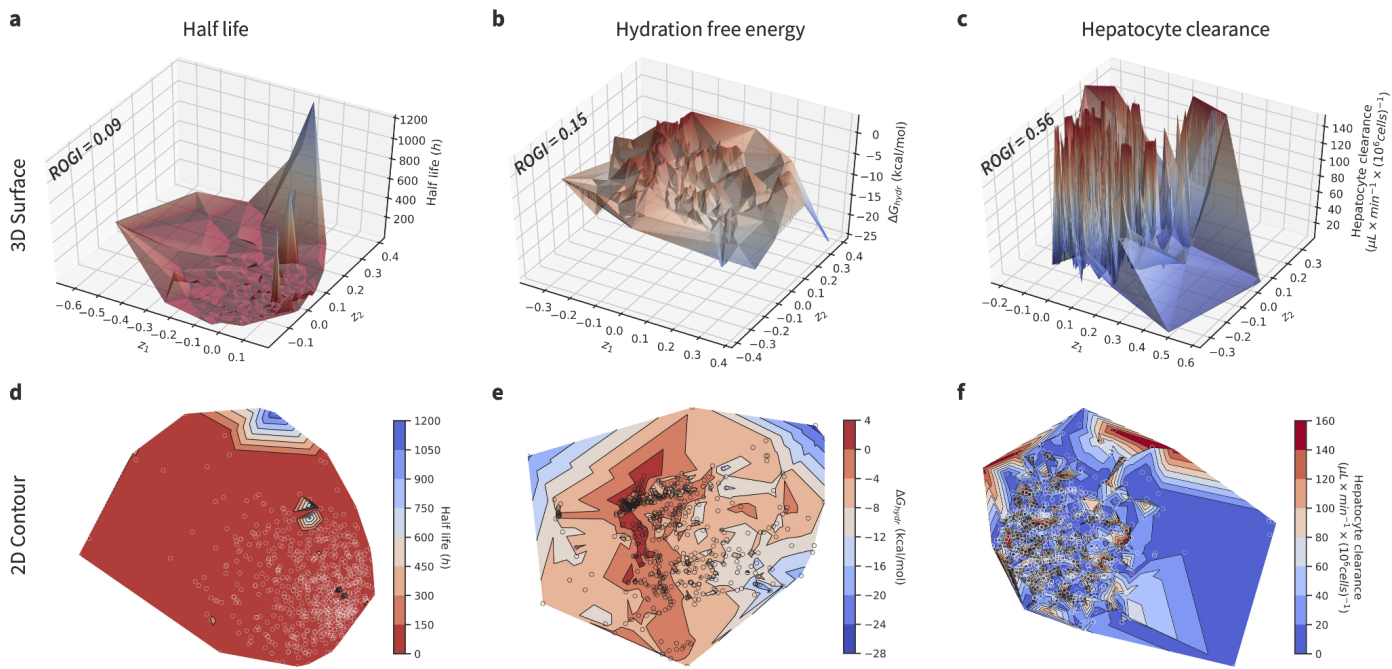

**Figure 6 | Visualization of molecular landscapes with different roughnesses.** Three high-dimensional datasets are projected onto the 2D plane (coordinates $z_1$ and $z_2$) by multidimensional scaling (MDS), with the third dimension being the property values. The landscapes are visualized as (a,b,c) three-dimensional surfaces, and (d,e,f) two-dimensional contour plots. The molecules in these datasets were described by a set of physico-chemical descriptors (Methods) and distances between them were computed as Euclidean distances. The ROGI for each dataset is shown. Qualitative differences between the roughness of these three datasets are visible and are in agreement with the computed ROGI.

## Visualizing structure-property landscapes

Visual inspection of structure-property landscapes can provide qualitative insight into its roughness. While many dimensionality reduction approaches have been used to project molecular datasets onto the 2D plane, multidimensional scaling (MDS) is a natural choice[19] when the size of the datasets are relatively small (up to a few thousand molecules) because it tries to preserve the proportionality of pairwise distances.

When large enough, differences in landscape roughness are evident by visual inspection. Figure 6 shows 3D landscapes for three different datasets available from the TDC with increasing roughness. In these plots, the first two dimensions are unitless coordinates used by MDS for the two-dimensional projection, while the third dimension is the property associated with each molecule. The ROGI value obtained for these datasets is in agreement with the degree of roughness that can be visually evaluated. Half life is the smoothest landscape among the three, with a ROGI of 0.09. Most of the landscape is relatively flat, with only one noticeable peak of high half life, and two minor ones. Hepatocyte clearance is instead the roughest landscape, with a ROGI of 0.56. Aside from a flat region of low clearance with low data density, the landscape is highly rugged with seemingly very similar molecules having very different clearance profiles. Finally, hydration free energy sits in the middle, with a ROGI of 0.15. This landscape does not have large regions being particularly flat or rough, but is somewhat rugged throughout. Three-dimensional landscapes for all the 12 TDC datasets studied here are shown in Figure S12.

## Discussion

Despite its quadratic scaling with dataset size, ROGI has a few advantages over modellability and SAR indices currently used. First, it is a general approach that may be applied to any structure-property relationships, rather than being confined to biological activity or requiring calibration. While it was primarily conceived for regression tasks, it may be applied to binary classification tasks too. Second, it more strongly correlates with model error than existing indices for regression, suggesting it better captures the roughness of molecular datasets by this measure. We reiterate that there is no ground-truth roughness value for molecular datasets to compare to. However, roughness is expected to relate to modellability. Hence we used model error in out-of-sample predictions as a way to quantitatively evaluate ROGI.

It is important to keep in mind that ROGI strongly depends on the molecular representation and the distance metric used. To some extent, this requirement is made necessary by the nature of chemical space, which does not have an inherent or obvious metric. This dependence may make comparing ROGI values obtained for different representations challenging. In particular, shifts in pairwise distance distributions between representations result in accompanying shifts in ROGI values (Figure S5). While in specific cases it may be possible to match distance distributions to make ROGI values for different representations comparable, a general solution may be elusive given how the notion of proximity is tightly linked to the definition of the metric being applied. Yet, it might be possible to use ROGI to compare variants of the same representation, such as different different types of fingerprints and different sets of physico-chemical descriptors. The concept of *intrinsic dimensionality*, the smallest number of variables needed to faithfully represent a dataset, is also dependent on the metric being applied and has been studied in the context of molecular simulations[80] and QSAR feature selection[81]. The main difference with ROGI is that the structure of the dataset is considered on its own and not in relation to a molecular property. However, further study on intrinsic dimensionality for molecular datasets may help quantify how the notion of proximity depends on the metric chosen and how it can affect ROGI values.

In practice, ROGI values are always estimates based on a finite sample of molecules. Therefore, ROGI also depends on the size of the dataset considered. In general, rougher landscapes are expected to require larger datasets for ROGI to converge to its true value (Figure S2). For the ZINC+GuacaMol datasets, 1000 molecules were sufficient to accurately estimate ROGI, while 100 were enough only for some properties (Figures S13 and S14). Analyzing the distribution of ROGI values for subsets of the data of various sizes is a simple way to assess whether the estimate may have converged. Note, however, this is a necessary but not sufficient condition to guarantee convergence. An approach to quantitatively assess convergence is to estimate the uncertainty of the ROGI estimate. While we do not provide a statistical estimator of ROGI uncertainty, a general approach to estimate uncertainty is by bootstrap[82]. Indeed, we find that bootstrap estimates of ROGI uncertainty correlate with its error, despite a general tendency to underestimate it (Figures S15 and S16).

Finally, ROGI is affected by noise in the molecular property for which roughness is being computed. Because ROGI does not model aleatoric noise, all property values are assumed to be exact. As such, noise (e.g., due to measurement error) introduces biases in the estimate based on roughness of the original landscape and the details of the noise. In general, ROGI is overestimated for smooth, especially flat, landscapes, while it is underestimated for rough landscapes (Figure S3). In specific fortuitous cases, in which the landscape displays a medium degree of roughness, or in which smooth and rough areas exist in different areas of the landscape, cancellation of error may lead to unbiased ROGI estimates despite the presence of noise. A possible approach to mitigate the effect of aleatoric noise could be to build a regression model that is able to take noise into account, and replace the measured property values with modeled ones.

Rather than evaluating correlation with regression error, it is natural to ask how the roughness of a molecular landscape affects molecular optimization performance. However, this is not a straightforward question to answer, as optimization introduces several considerations and confounding variables. First, optimization performance is expected to depend on the optimization strategy adopted (e.g., gradient, evolutionary, model-based) more strongly than regression performance does on the ML model chosen. Second, contrary to regression, optimization difficulty depends on the *distribution* of property values in an asymmetric fashion. Specifically, optimization difficulty depends on how skewed a property distribution is toward the optimal/preferred values. A "needle in a haystack" scenario is challenging partly because there are only a few solutions among many available options. If the situation were reversed (e.g., very few bad molecules, and mostly good ones), the optimization task would become trivial. Roughness and regression difficulty are not affected by whether the distribution of property values is skewed toward larger or smaller values. Finally, multiple performance measures for optimization are available, each giving more weight to different aspects of the optimization depending on the application (e.g., best observed value, sample efficiency), such that a universally accepted measure of performance is harder to establish. For these reasons, the development of a quantitative measure of molecular optimization difficulty requires substantial modifications to ROGI and is thus left for future work. One possibility may be to consider higher, odd-order moments of the property distribution rather than its variance.

## Conclusion

We have developed and presented ROGI, a quantitative measure of structure-property landscape roughness for molecular datasets that may be used for exploratory data analysis in molecular design campaigns. ROGI is applicable to any structure-property relationship for which property values and pairwise distances between molecules are available. We have tested the ability of ROGI to correlate with the error of regression ML models on 55 datasets covering a broad range of molecular properties, and have generally found strong correlations between ROGI values and model errors. For regression tasks, ROGI correlated to model error better than existing indices of landscape modellability and roughness. For binary classification tasks, both ROGI and existing indices proved valuable depending on the degree of dataset balance and whether accuracy or balanced accuracy is the performance measure of interest. Future work will focus on expanding ROGI to multi-label classification, and to molecular optimization. Finally, we note that ROGI can be applied to (bio)chemical systems beyond small organic molecules, such as macromolecules and crystalline materials.

## Acknowledgments
The authors would like to thank Wendy D. Cornell, Wenhao Gao, Rocío Mercado, and Thjis Stuyver for valuable discussions and suggestions. This work was funded by the MIT-IBM Watson AI Lab.

## Data and software availability

An implementation of ROGI is provided as a Python package at https://github.com/coleygroup/rogi. The package also includes implementations of SARI, MODI, and RMODI. Results and associated analysis can be reproduced using the data and Jupyter Notebooks available at https://github.com/coleygroup/rogi-results.

**Supplementary Information**

# Roughness of Molecular Property Landscapes and Its Impact on Modellability

Matteo Aldeghi [1], David E. Graff [1,2], Nathan Frey [3], Joseph A. Morrone [4], Edward O. Pyzer-Knapp [5], Kirk E. Jordan [6], Connor W. Coley [1,7,*]

[1] Department of Chemical Engineering, Massachusetts Institute of Technology, Cambridge, MA 02139, USA
[2] Department of Chemistry and Chemical Biology, Harvard University, Cambridge, MA 02138, USA
[3] Lincoln Laboratory, Massachusetts Institute of Technology, Lexington, MA 02421, USA
[4] IBM Thomas J. Watson Research Center, Yorktown Heights, NY 10598, USA
[5] IBM Research Europe, Daresbury, UK
[6] IBM Thomas J. Watson Research Center, Cambridge, MA 02142, USA
[7] Department of Electrical Engineering and Computer Science, Massachusetts Institute of Technology, Cambridge, MA 02139, USA
[*] Email: ccoley@mit.edu

**Figure S1 | Examples of extreme ROGI values.** (a) A dataset with the same property value across all molecules has ROGI of zero. (b) A dataset in which all nearest neighbors (here so close to appear on top of each other) have opposite property values has ROGI of one. (c-d) As distance between molecules with opposing property values increases, ROGI decreases.

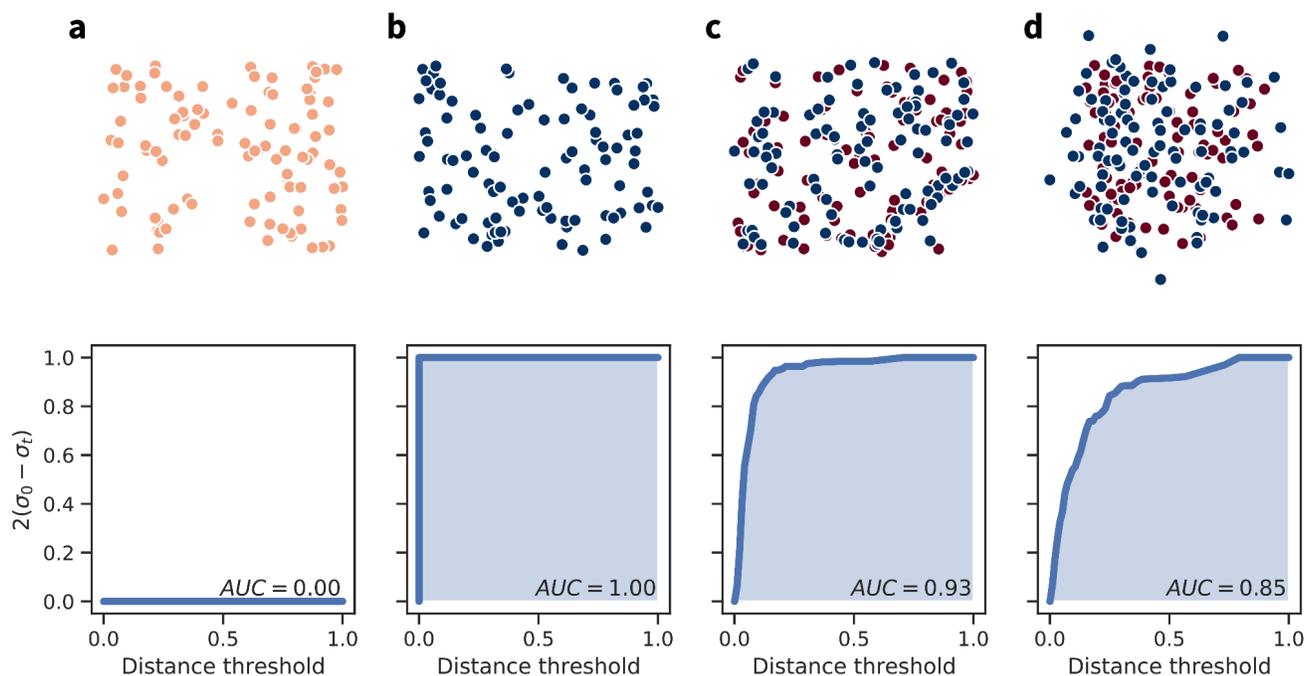

**Figure S2 | Convergence of ROGI values with respect to dataset size on toy surfaces.** By necessity, ROGI is an estimate based on a finite sample of chemical space. Depending on the details of the structure-property landscape, ROGI might converge faster or slower toward its true value with increased sample size. Landscapes like those of S4 are particularly susceptible to small sample sizes due to having small areas with property values very different from the average ones. Rough surfaces, like S5 and S6, also require larger sample sizes to capture the presence of a large number of sharp cliffs. Smoother landscapes (S1, S2) and those with a more uniform distribution of property values (S3) are generally expected to display faster convergence. Unfortunately, the convergence properties of ROGI cannot be known a priori for any given structure-property landscape, as it depends on the landscape itself, which is unknown.

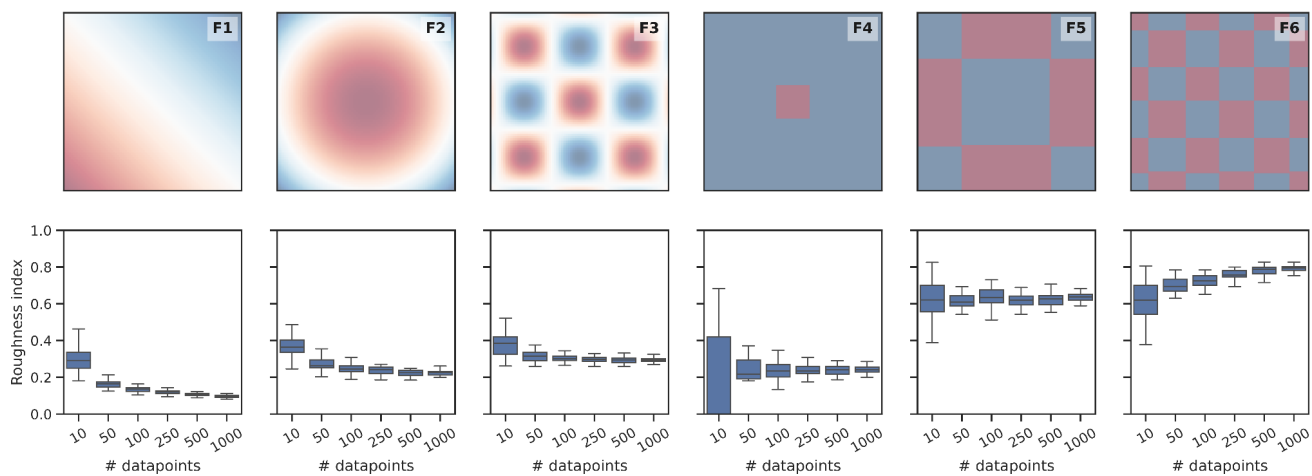

**Figure S3 | Effect of noise on ROGI estimates.** Here, we added Gaussian noise to the property values of datasets with 100 samples, which were constructed as shown in Figure 2b. Different levels of noise, relative to the range or property values observable in S1–S6, was added. For each noise level, this procedure was repeated 50 times. (a) Noiseless surfaces. (b) Surfaces with noise added. Many property values exceed the bounds observed in the noiseless setting. (c) Surfaces with noise added and renormalized. Due to the normalization, the distribution of property values is shifted toward average (white) values. However, noise also disrupts the smooth relationship between neighboring points on the surface. (d) Mean and standard deviation of the ROGI values obtained for 50 noisy datasets. A red, dashed horizontal line shows the ROGI value for the noiseless dataset as a reference. These results show that when noise is present, ROGI tends to be overestimated for flat landscapes, and underestimated for rough landscapes. Without normalization based on the range of property values, ROGI can exceed 1. ROGI values computed without normalization for the noisy surface in (b) would always be higher than those for their noiseless counterparts.

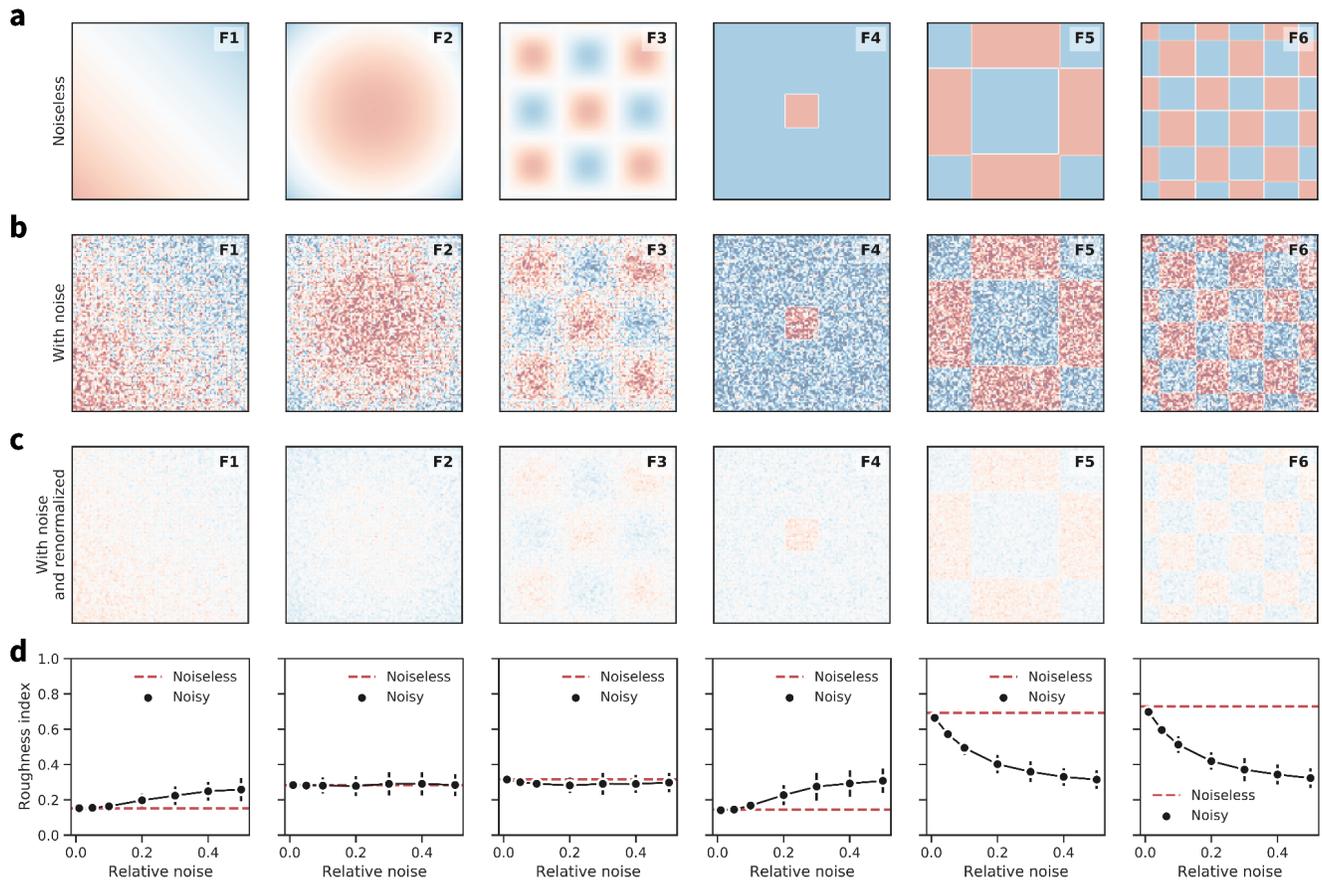

**Figure S4 | Distribution of cross-validation RMSEs by model, representation, and dataset group. A smaller range of** RMSE values was observed for ChEMBL with respect to ZINC+GuacaMol and TDC datasets. The range of ROGI values is also tighter for ChEMBL.

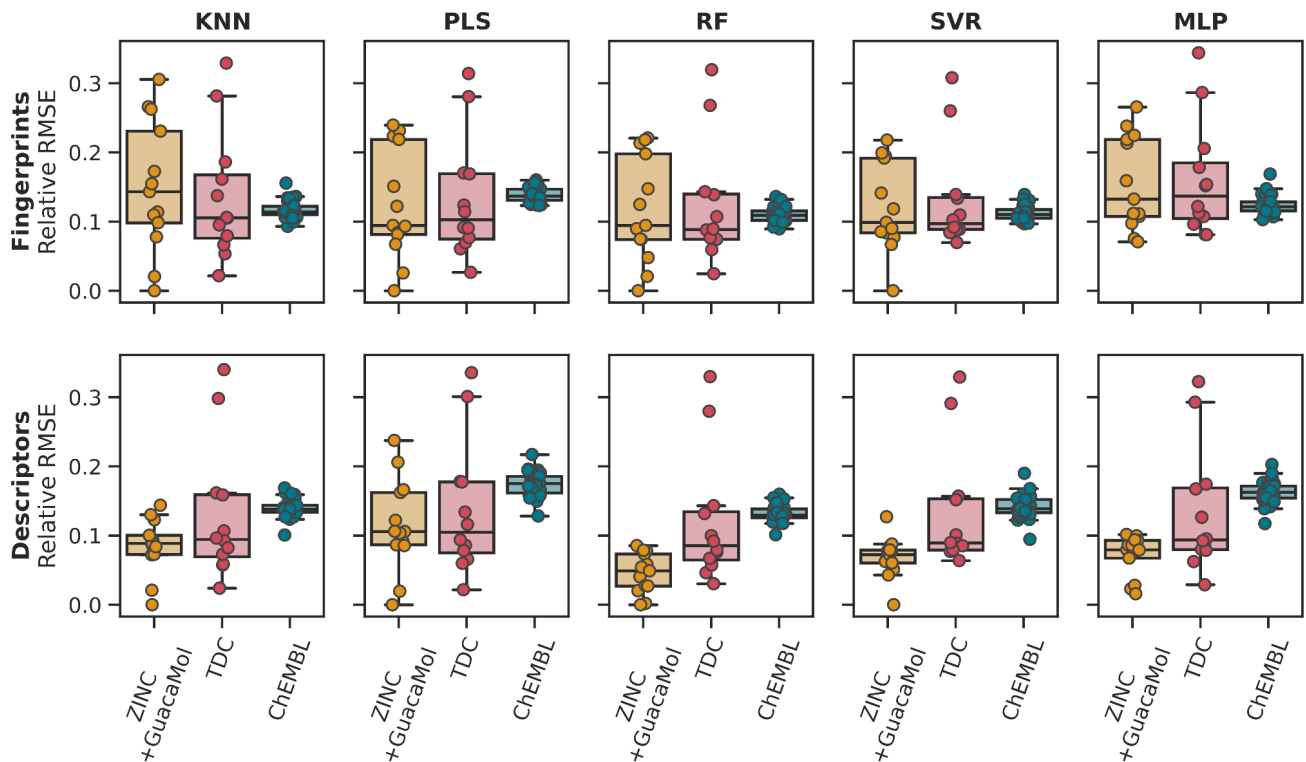

**Figure S5 | Distributions of pairiwse Tanimoto (fingerprints) and Euclidean (descriptors) distances for the molecules considered in the ZINC+GuacaMol datasets.** Shown are the kernel density estimates for 100,000 pairwise distances sampled at random from the set of ~2 million distances. Tanitomo distances based on fingerprint representations are more tightly distributed and are shifted towards higher values, compared to Euclidean distances based on descriptors. The different distance distributions affect the ROGI values obtained with these two different representations, as molecules described via physico-chemical descriptors will appear closer to each other than those described via fingerprints.

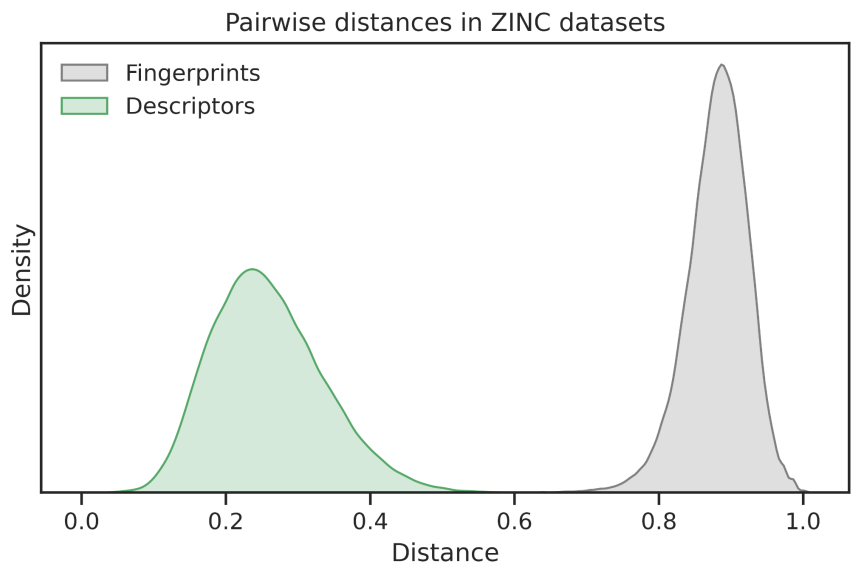

**Figure S6 | Correlation between SARI and cross-validated model error for the set of 30 ChEML regression tasks.** Results for ROGI are shown as a comparison. Here, we also considered MACCS keys as a representation, as it was the representation used in the original SARI publication[13]. Overall, ROGI achieved higher linear correlations with cross-validated model errors than SARI.

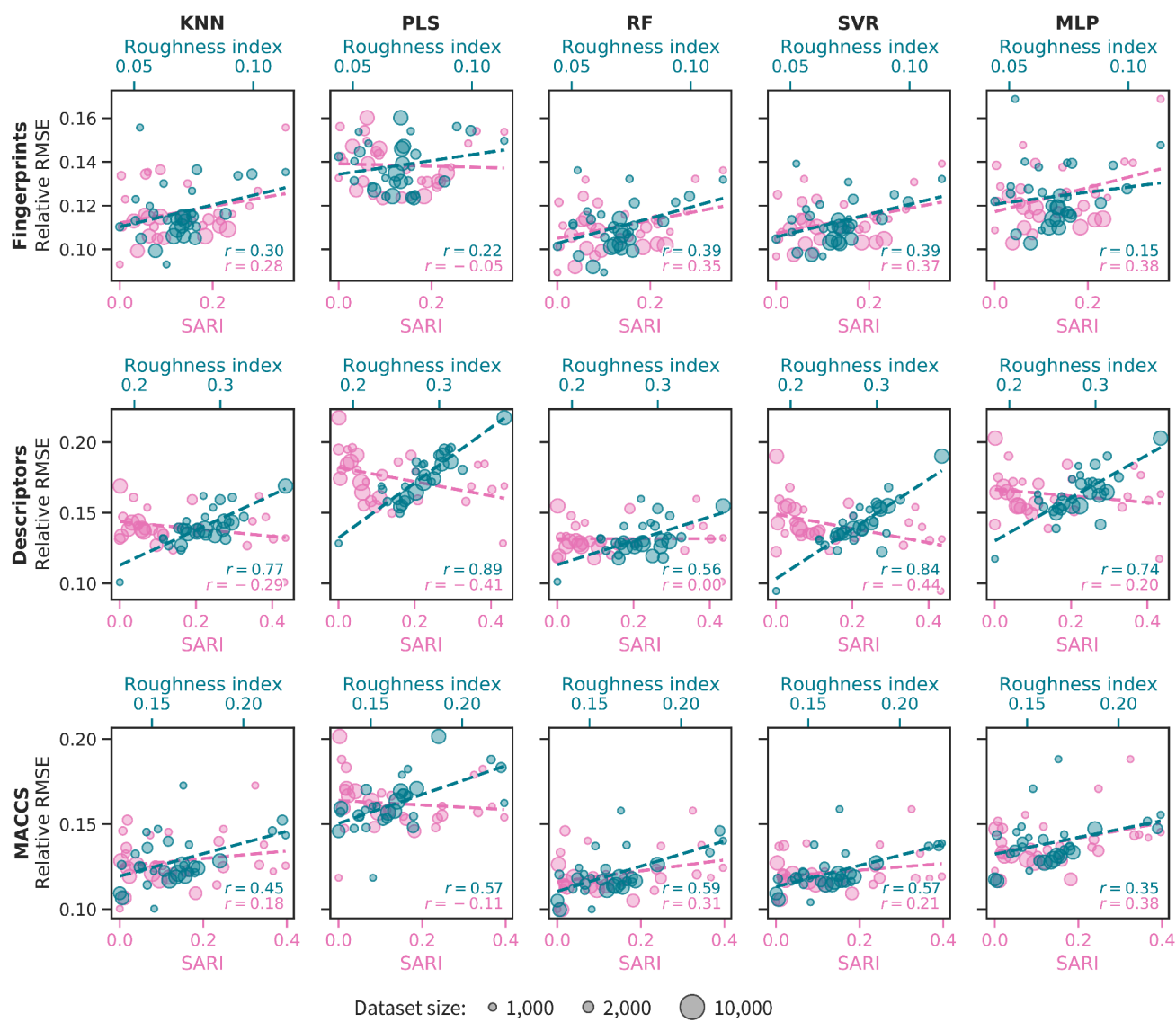

**Figure S7 | Correlation between RMODI and cross-validated model error for a set of 55 regression tasks.**
The relative RMSE is the average, normalized RMSE obtained from 5-fold cross validation. These results may be compared to those obtained with ROGI and shown in Figure 4.

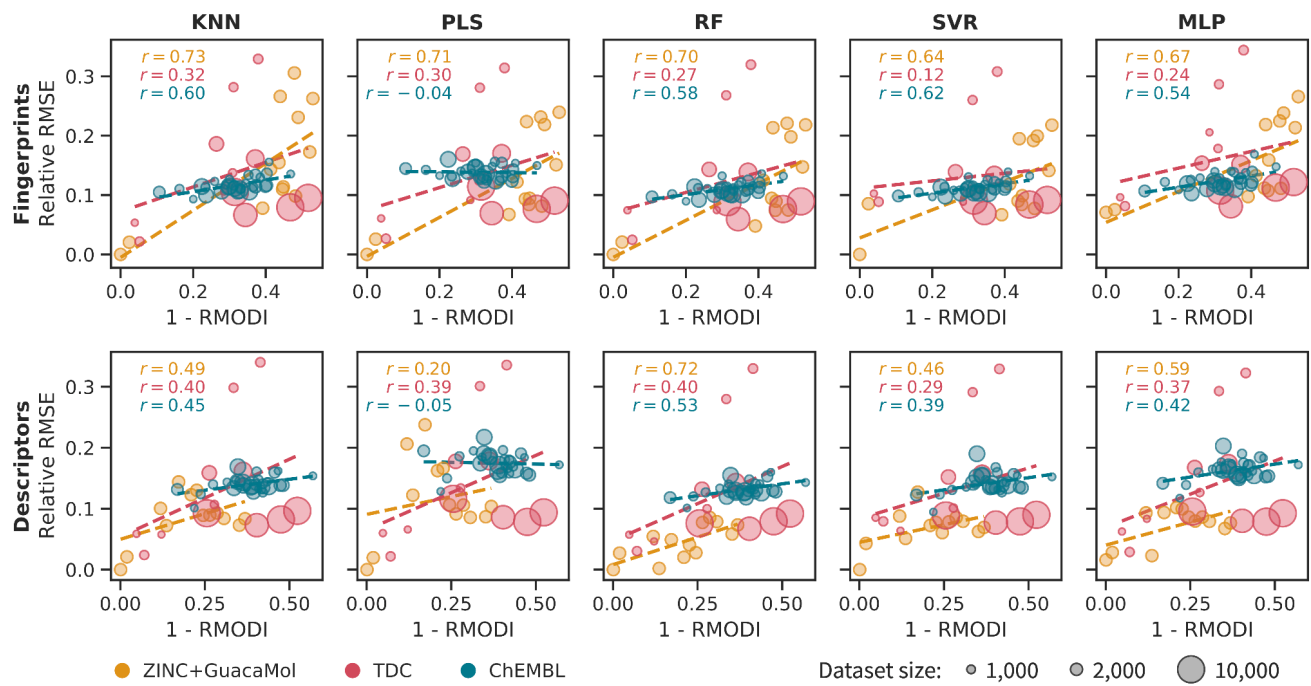

**Figure S8 | Correlation between the MODI and cross-validated binary accuracy for a set of 50 classification tasks.** These results may be compared to those obtained with ROGI and shown in Figure 5.

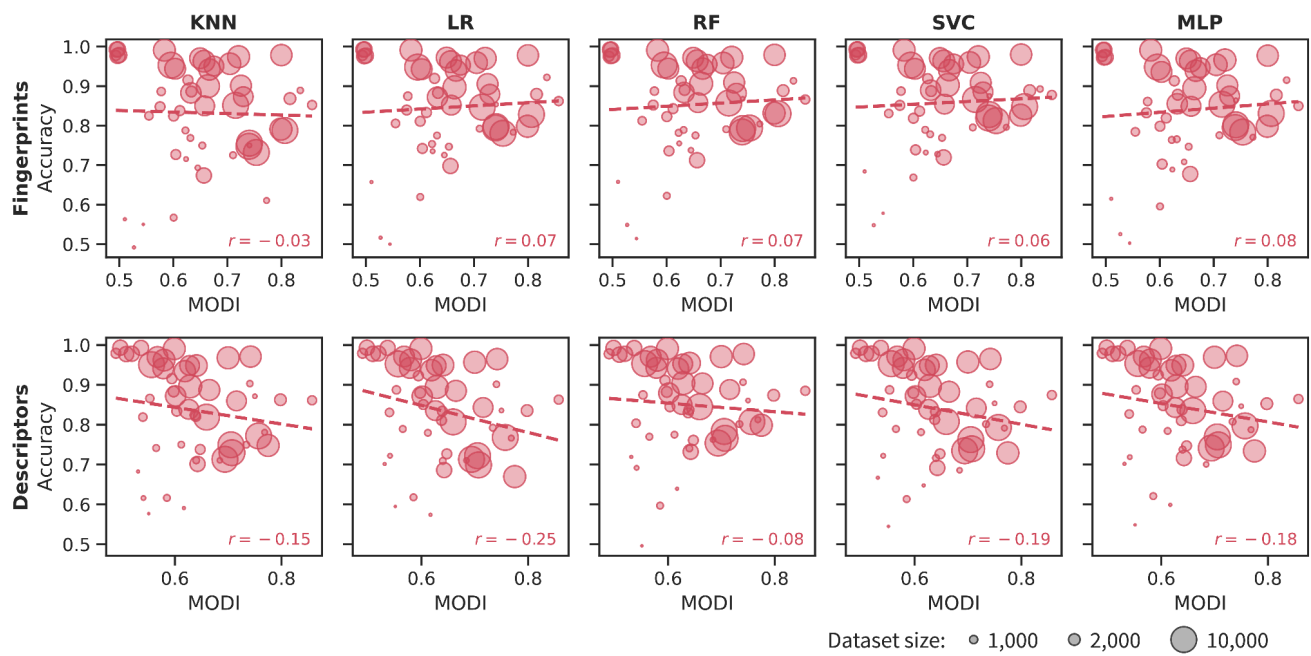

**Figure S9 | Correlation between the MODI and cross-validated balanced accuracy for a set of 50 classification tasks.** These results may be compared to those obtained with ROGI and shown in Figure S10.

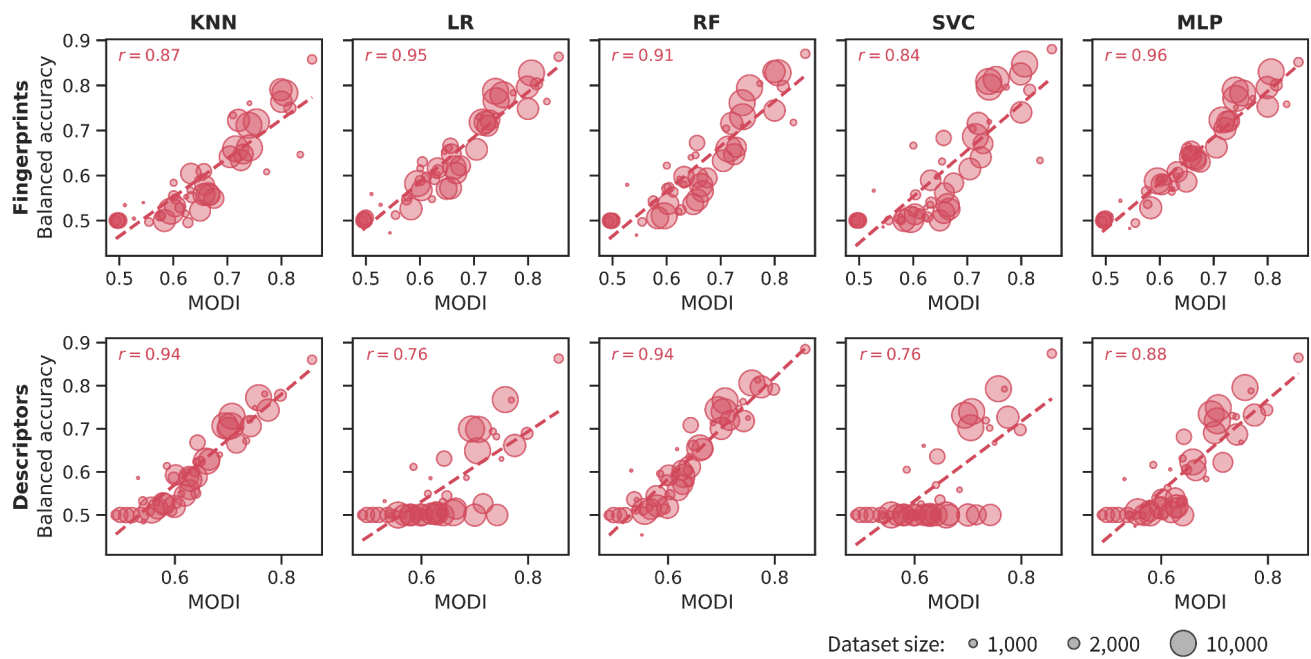

**Figure S10 | Correlation between the ROGI and cross-validated balanced accuracy for a set of 50 classification tasks.** These results may be compared to those obtained with MODI and shown in Figure S9.

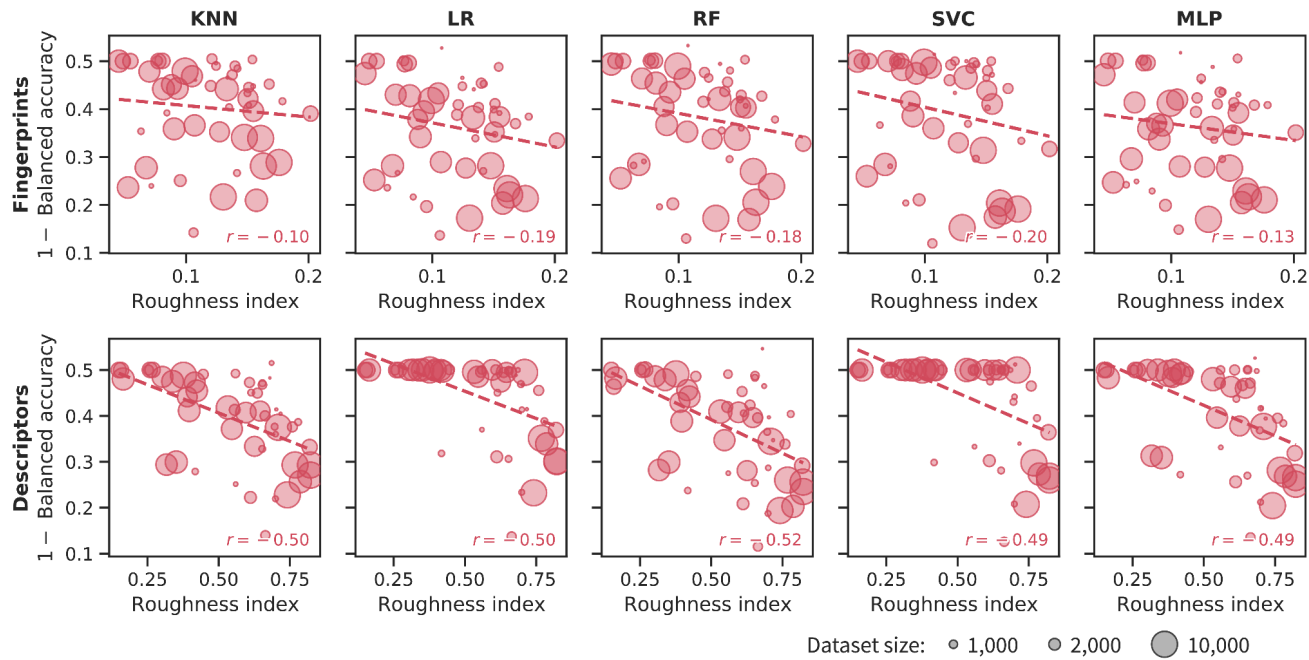

**Figure S11 | Correlation between the ROGI / MODI and cross-validated binary accuracy for a set of 55 classification tasks with perfect balance.** These datasets were artificially created by binarizing the regression datasets (Methods) such that half of the molecules had positive and half negative labels. In this way, ROGI and MODI can more directly be compared using binary accuracy. While both indices returned good correlations between their values and model error, MODI systematically provided higher correlations. ROGI returned moderate-to-strong linear correlations with binary accuracy, between 0.56 and 0.93. MODI returned stronger correlations, between 0.75 and 0.99.

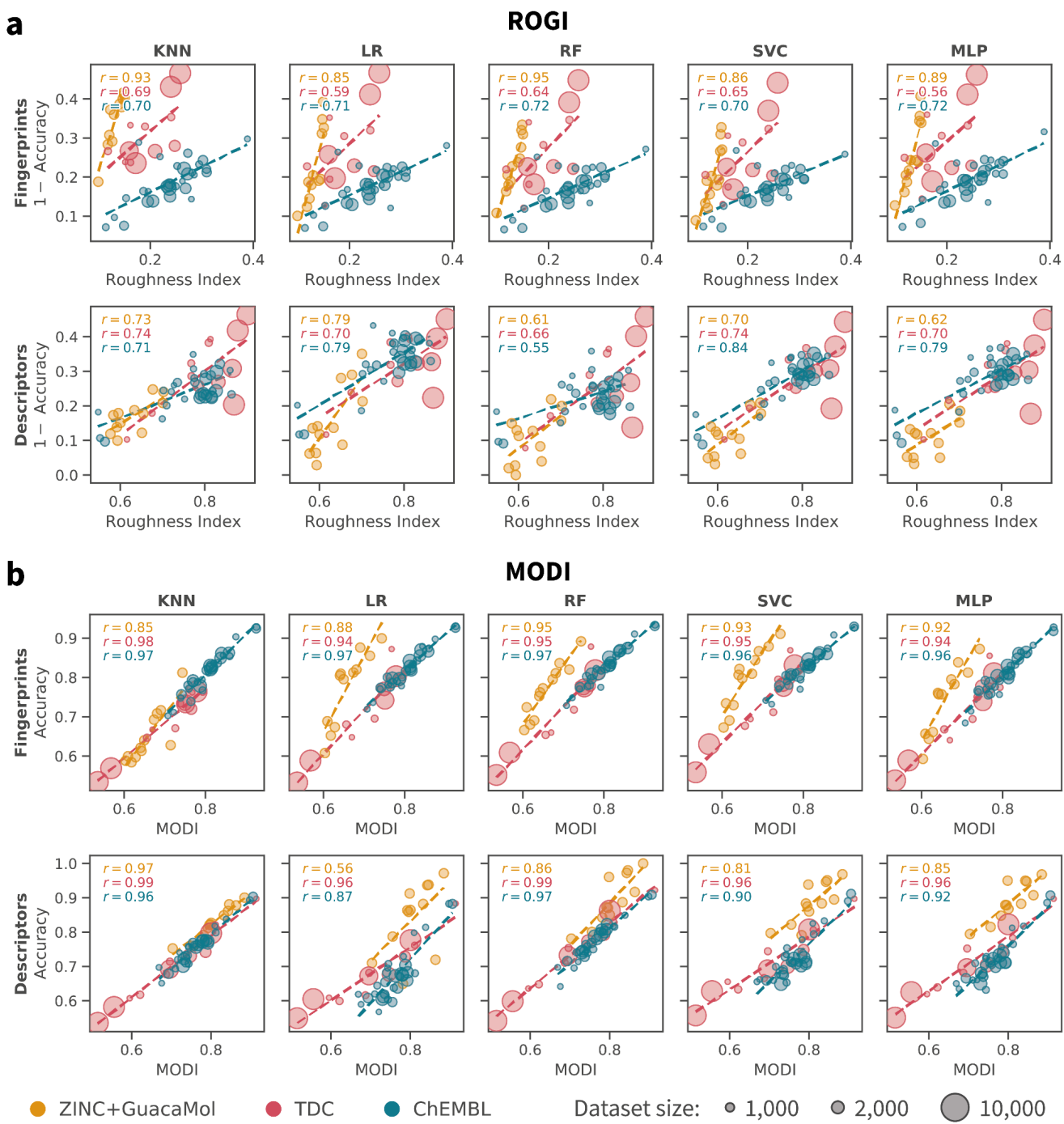

**Figure S12 | Visualization of molecular landscapes with different roughnesses.** Three high-dimensional datasets are projected onto the 2D plane (coordinates $z_1$ and $z_2$) by multidimensional scaling (MDS), with the third dimension being the property values. The landscapes are visualized as three-dimensional surfaces and two-dimensional contour plots. The molecules in these datasets were described by a set of physico-chemical descriptors (Methods) and distances between them were computed as Euclidean distances. The ROGI for each dataset is shown.

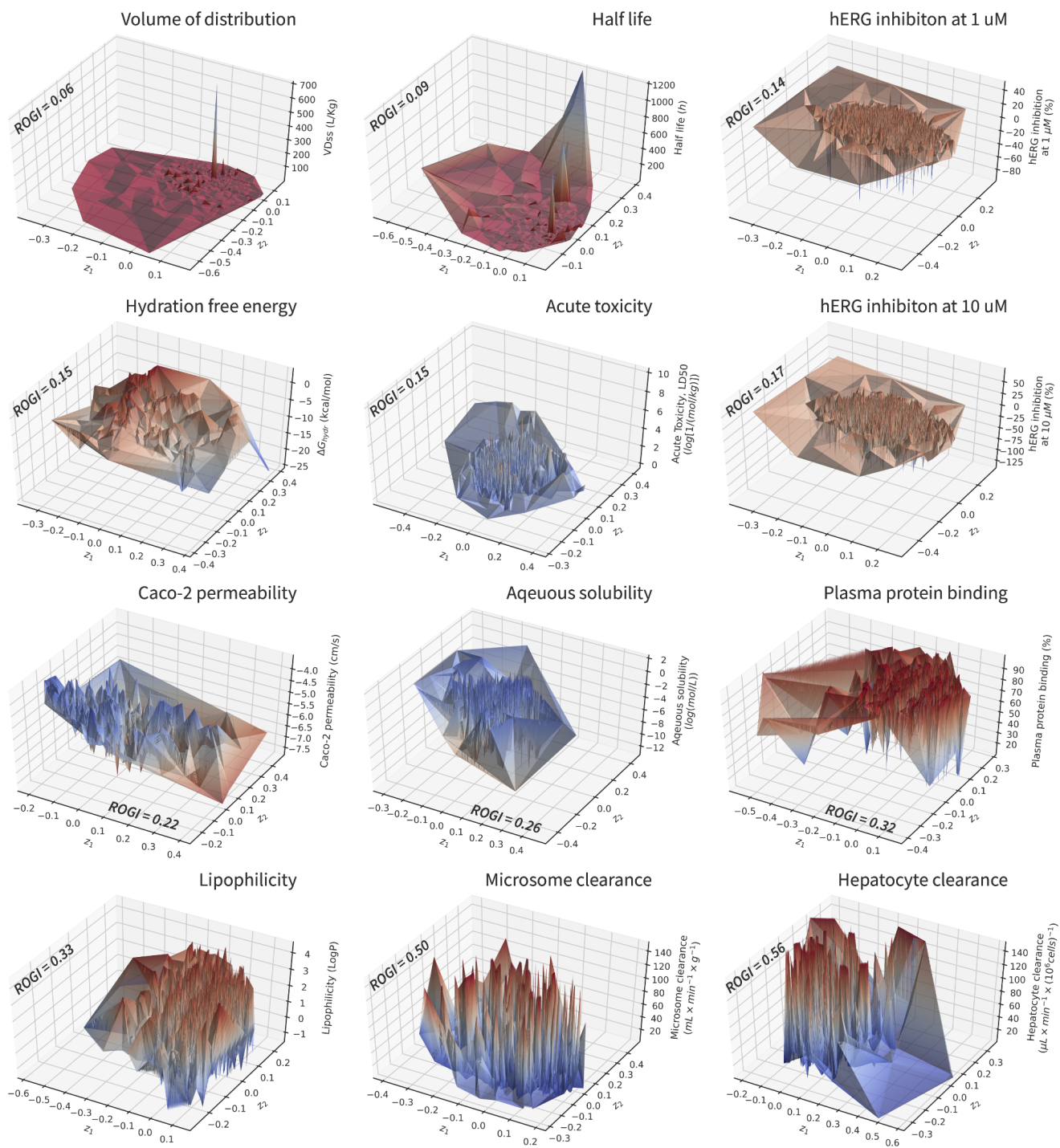

**Figure S13 | Convergence of ROGI values of ZINC+GuacaMol datasets of increasing size and fingerprint representations.** Shown are the distributions of ROGI values obtained for 10 random subsets of size 10, 100, 1000, 2000, 5000 of a dataset for 10,000 molecules sampled from ZINC. A black, horizontal dashed line indicates the ROGI value for the whole set of 10,000 molecules. These results were obtained for molecules described by binary Morgan fingerprints (Methods). Results for *Valsartan_SMARTS* are not shown, because both property values and ROGI are always zero (i.e., completely flat landscape for which ROGI immediately converges to zero). In general, subsets with at least 100–1000 molecules were needed for ROGI values to be close to the reference value obtained with 10,000 molecules.

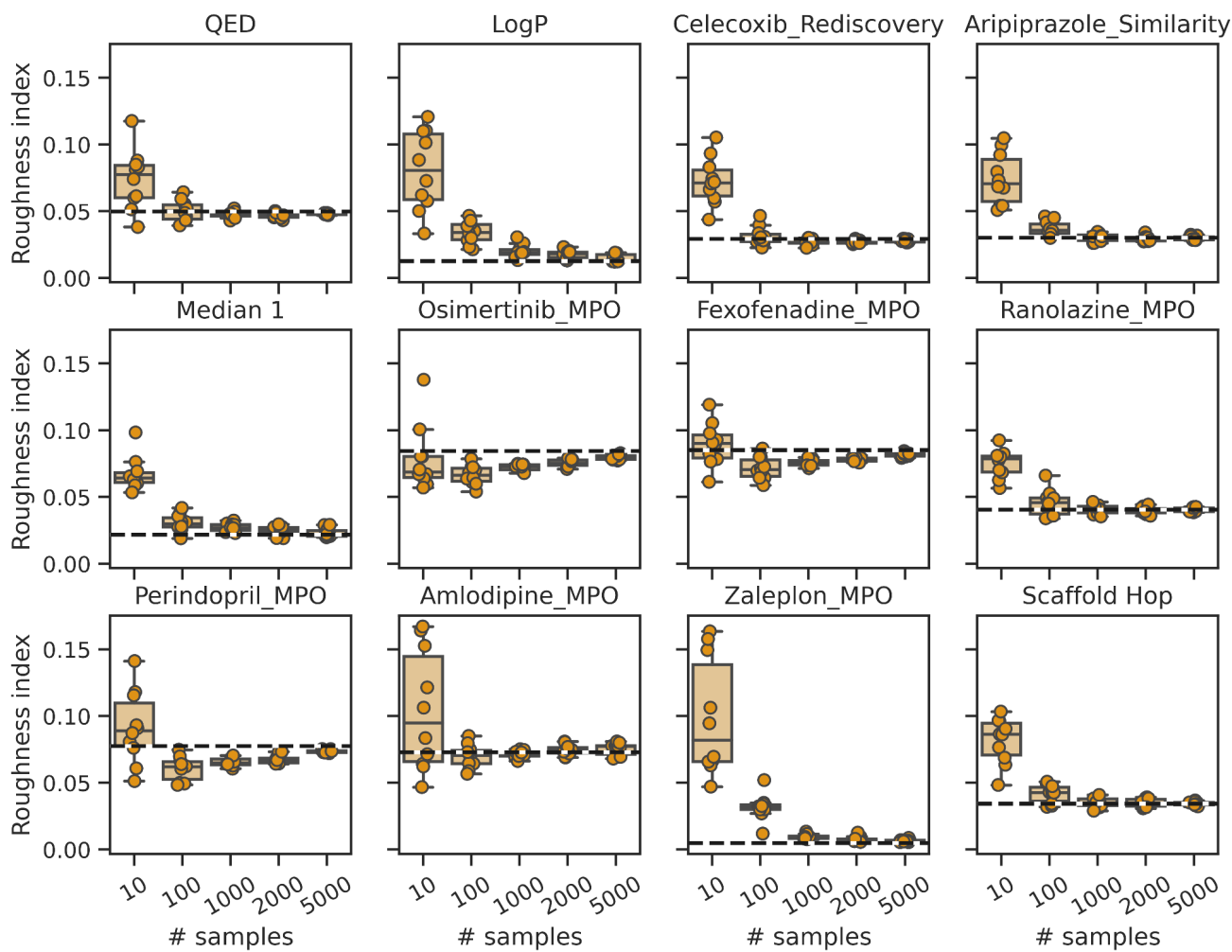

**Figure S14 | Convergence of ROGI values of ZINC+GuacaMol datasets of increasing size and descriptor representations.** Shown are the distributions of ROGI values obtained for 10 random subsets of size 10, 100, 1000, 2000, 5000 of a dataset for 10,000 molecules sampled from ZINC. A black, horizontal dashed line indicates the ROGI value for the whole set of 10,000 molecules. These results were obtained for molecules described by a set of physico-chemical descriptors (Methods). Results for *Valsartan_SMARTS* are not shown, because both property values and ROGI are always zero (i.e., completely flat landscape for which ROGI immediately converges to zero). In general, subsets of size >1000 were needed for ROGI values to be close to the reference value obtained with 10,000 molecules.

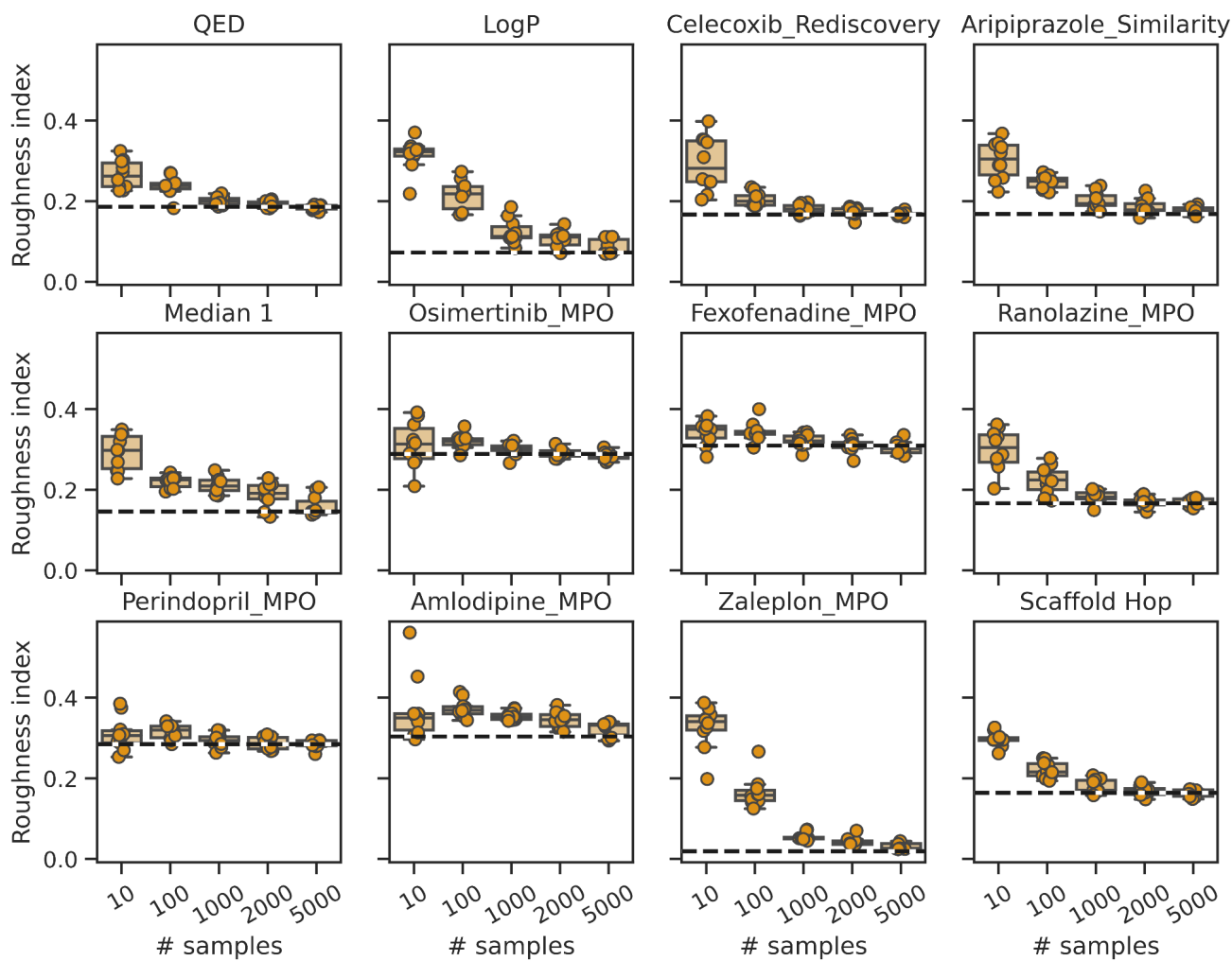

**Figure S15 | ROGI uncertainty estimation by bootstrap for ZINC+GuacaMol datasets and fingerprint representations.** 10 subsets of size 10, 100, 1000, 2000, 5000 of a dataset for 10,000 molecules sampled from ZINC+GuacaMol were considered. For each of these 50 subsets, ROGI, its uncertainty (computed with 20 bootstrap samples), and its error with respect to ROGI for the full dataset of 10,000 molecules, were computed. The parity plots compare the ROGI error on the x-axis to the size of the bootstrapped 95% confidence interval (CI) on the y-axis. These results were obtained for molecules described by binary Morgan fingerprints (Methods). Results for *Valsartan_SMARTS* are not shown, because both property values and ROGI are always zero (i.e., completely flat landscape for which ROGI immediately converges to zero). The size of the 95% CI generally correlates with, but also underestimates, the ROGI error.

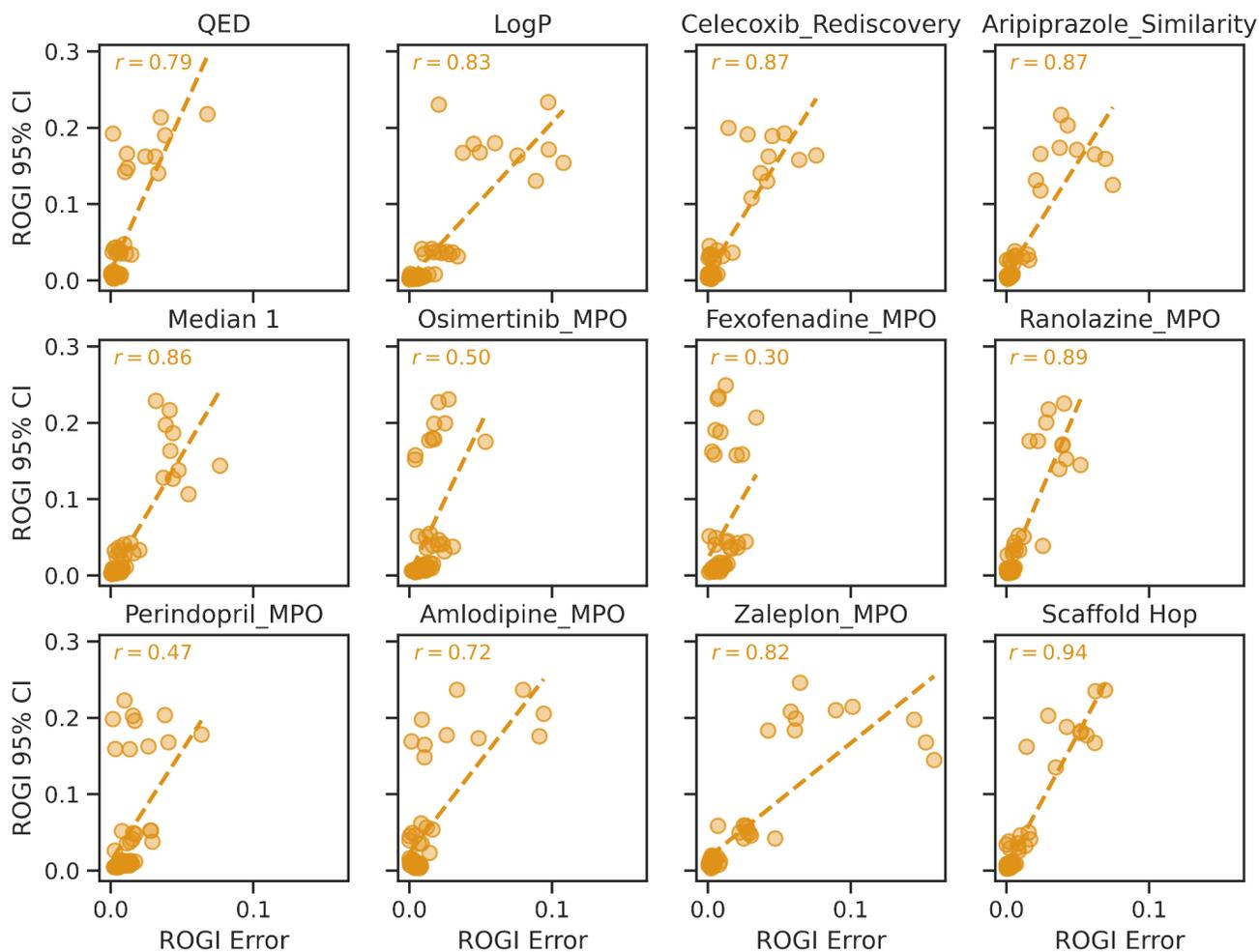

**Figure S16 | ROGI uncertainty estimation by bootstrap for ZINC+GuacaMol datasets and descriptor representations.** 10 subsets of size 10, 100, 1000, 2000, 5000 of a dataset for 10,000 molecules sampled from ZINC+GuacaMol were considered. For each of these 50 subsets, ROGI, its uncertainty (computed with 20 bootstrap samples), and its error with respect to ROGI for the full dataset of 10,000 molecules, were computed. The parity plots compare the ROGI error on the x-axis to the size of the bootstrapped 95% confidence interval (CI) on the y-axis. These results were obtained for molecules described by a set of physico-chemical descriptors (Methods). Results for *Valsartan_SMARTS* are not shown, because both property values and ROGI are always zero (i.e., completely flat landscape for which ROGI immediately converges to zero). The size of the 95% CI generally correlates with, but also underestimates, the ROGI error.

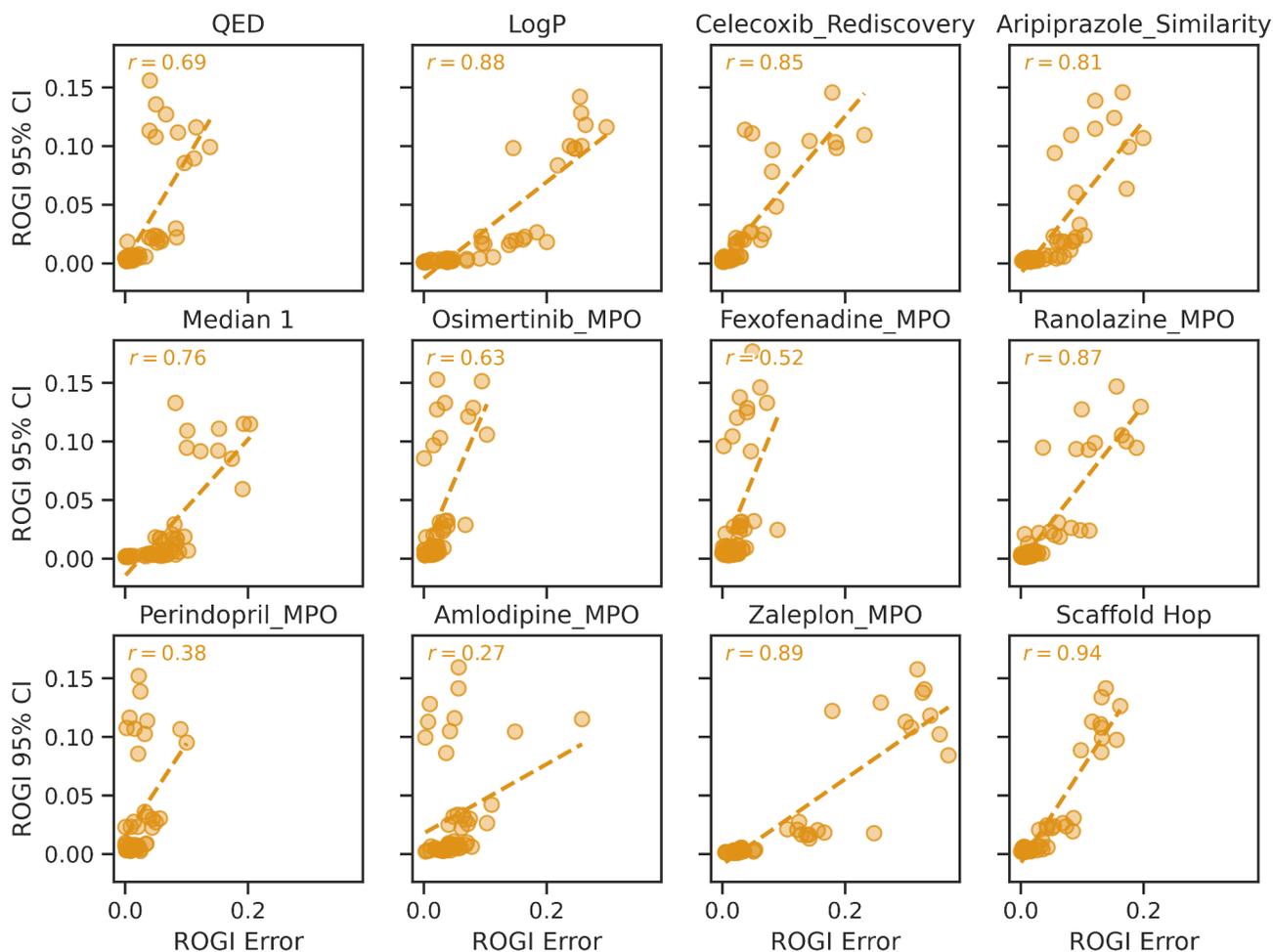